\documentclass[12pt,tightenlines]{revtex4} 
\usepackage{amssymb,amsmath}
\usepackage{psfrag}
\usepackage[dvips]{graphicx}
\usepackage{epic,curves}%,verbatim}

\allowdisplaybreaks
\newcommand{\mean}[1]{\langle#1\rangle}
\newcommand{\Det}{\mathop{\rm Det}\nolimits}
\newcommand{\logequiv}{\ \Leftrightarrow\ }

\newcommand{\intOinfty}{\displaystyle\int_{0}^{\infty}}
\newcommand{\hc}{\mathrm{h.c.}}

\begin{document}

\title{To the practical design of the optical lever intracavity topology of
gravitational-wave detectors}

\author{S.L.Danilishin}
\email{stefan@hbar.phys.msu.ru}
\affiliation{Physics Faculty, Moscow State University, Moscow 119992, Russia}

\author{F.Ya.Khalili}
\email{farid@hbar.phys.msu.ru}
\affiliation{Physics Faculty, Moscow State University, Moscow 119992, Russia}

%\date{Draft of \today}

\begin{abstract}

The QND intracavity topologies of gravitational-wave detectors proposed
several years ago allow, in principle, to obtain sensitivity significantly 
better than the Standard Quantum Limit using relatively small anount of 
optical pumping power. In this article we consider an improved more
``practical'' version of the {\it optical lever} intracavity scheme. It
differs from the original version by the symmetry which allows to suppress
influence of the input light amplitude fluctuation. In addition, it provides
the means to inject optical pumping inside the scheme without increase
of optical losses.
  
We consider also sensitivity limitations imposed by the {\it local meter} 
which is the key element of the intracavity topologies. Two variants of the
local meter are analyzed, which are based on the spectral variation
measurement and on the Discrete Sampling Variation Measurement, 
correspondingly.  The former one, while can not be considered as a candidate
for a practical implementation, allows, in  principle, to obtain the best
sensitivity and thus can be considered as an  ideal ``asymptotic case'' for
all other schemes. The DSVM-based local meter can be considered as a realistic
scheme but its sensitivity, unfortunately, is by far not so good just due to a
couple of peculiar numeric factors specific for this scheme.

From our point of view search of new methods of mechanical QND  measurements
probably based on improved DSVM scheme or which combine the local meter with
the pondermotive squeezing technique, is necessary.

\end{abstract}

\maketitle

%\tableofcontents

\section{Introduction}

The large-scale laser interferometric gravitational-wave detectors 
\cite{Abramovici1992, Caron1997, Ando2001, Willke2002} which has been built to
search gravitational waves from very distant astrophysical sources represent
now the most sensitive measurement devices for mechanical acceleration and
displacement. Currently their sensitivity is close to
$\sqrt{S_x}=10^{-19}\,{\rm m/Hz^{1/2}}$ in frequency range  $100\div 200\,{\rm
Hz}$ \cite{Whitcomb2005}. This value is only $\sim 30$ times larger than the
Standard Quantum Limit (SQL) of these devices sensitivity
\cite{67a1eBr, 92BookBrKh, 03a1BrGoKhMaThVy}.

The next generation of terrestrial gravitational-wave detectors probably will
reach this limit in 2008-2010 \cite{WhitePaper1999, Fritschel2002}, and then
overcome it. The overcoming of the SQL will require more or less significant
modification  of the detectors topology. Several variants of this modification
have been  proposed. They can be divided into two groups. 

The first group  
\cite{
02a1KiLeMaThVy, 
Buonanno2001, 
01a2Kh,  
Buonanno2002,  
Buonanno2003,   
Harms2003,
Buonanno2004,
00a1BrGoKhTh, 
Purdue2001,  
Purdue2002,   
Chen2002,  
02a2Kh, 
04a1Da} 
(which can be considered as the ``mainstream'') preserves in general the current
detector topology. We will refer to these schemes below as {\it extracavity} 
ones because all of them convert phase shift of the optical pumping field 
created by the gravitational-wave signal into some modulation of the output 
light beam which is detected by photodetector(s) {\it outside} the
interferometer optical cavities.

Unfortunately, due to semi-technological limitations common for all these
schemes \cite{00p1BrGoKhTh} they can not provide sensitivity significantly
better than the SQL. The second group of methods, so-called {\it intracavity}
schemes \cite{96a2BrKh, 97a1BrGoKh, 98a1BrGoKh, 02a1Kh, 03a1Kh}, requires more
radical modification of the detector topology but can provide substantially
better sensitivity with smaller value of optical pumping power. The  basic idea of this
method was proposed in the article \cite{96a2BrKh} and can  be formulated as
the following: measure directly the redistribution of optical
energy created by the gravitational wave {\em inside} the detector in a QND way
(without absorption of optical quanta).

\begin{figure*}[t]
  \psfrag{E1}[cb][lb]{${\sf E}_1$}
  \psfrag{E2}[rc][lb]{${\sf E}_2$}
  \psfrag{I1}[cb][lb]{${\sf I}_1$}
  \psfrag{I2}[rc][lb]{${\sf I}_2$}
  \psfrag{C}[lb][lb]{{\sf C}} \psfrag{D}[rt][rt]{{\sf D}}
  \includegraphics[width=0.45\textwidth]{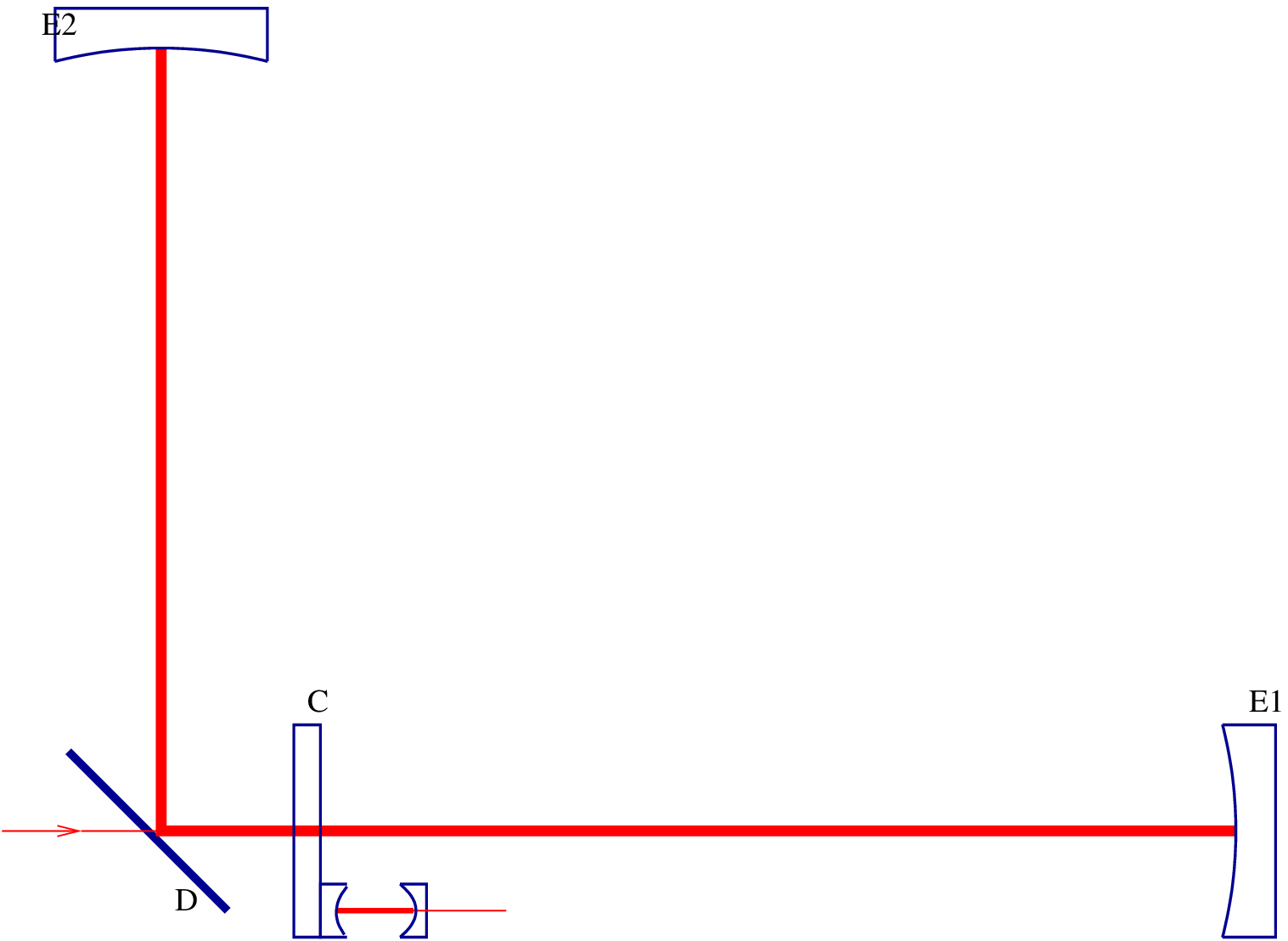}\hspace{\fill}
  \includegraphics[width=0.45\textwidth]{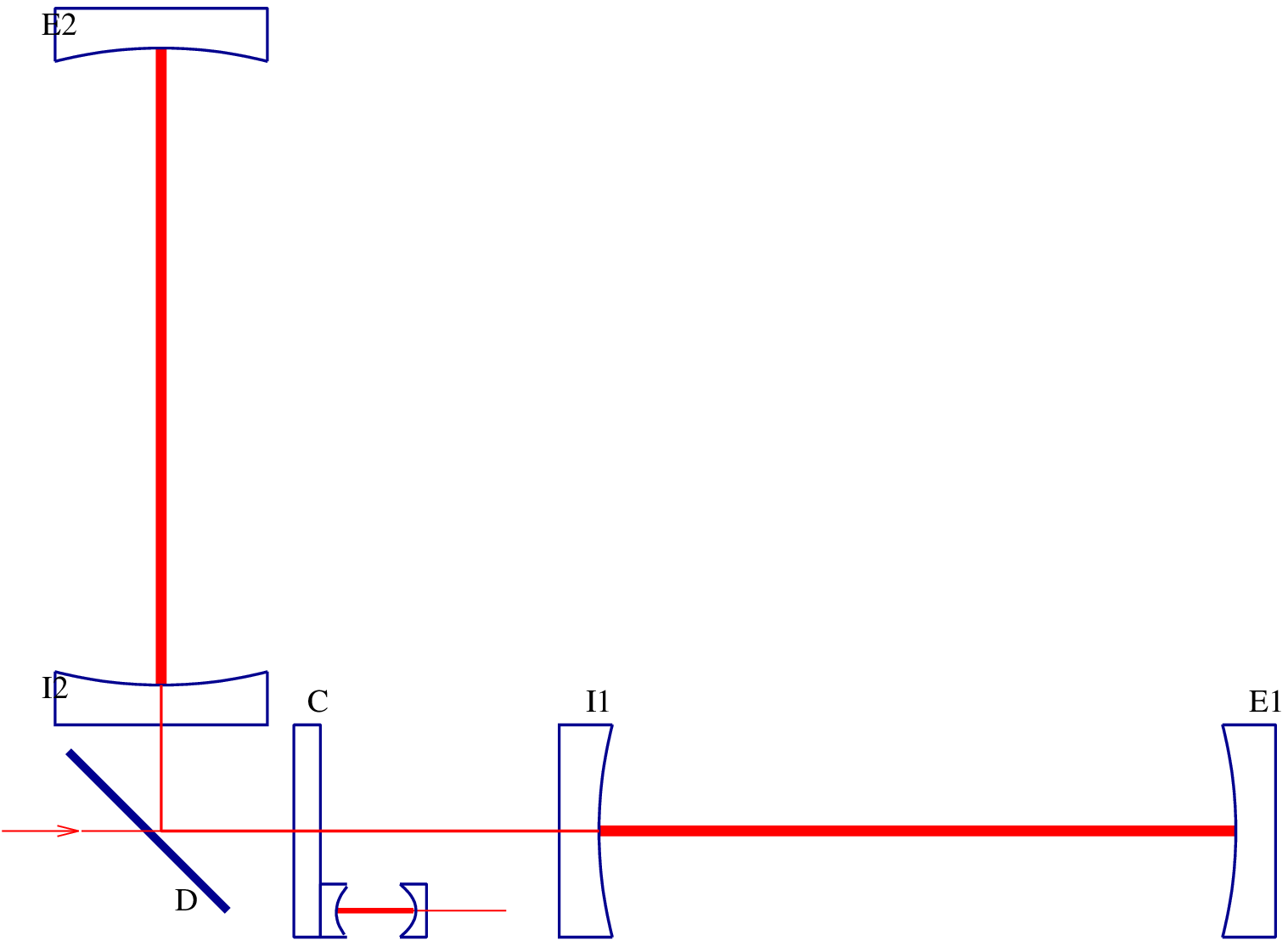}
  \caption{The ``optical bars'' (left) and ``optical lever'' (right)
  intracavity schemes}\label{fig:optbar}
\end{figure*}

In the article \cite{97a1BrGoKh} possible implementation of this idea, the
{\it optical bars} scheme was proposed (see Fig.\,\ref{fig:optbar}, left). In
this scheme the end mirrors ${\sf E}_1$, ${\sf E}_2$ and the central mirror
{\sf C} form two Fabry-Perot cavities coupled by means of a partly
transparent mirror {\sf C}. Relatively weak external optical pumping is
necessary in order to compensate internal losses in the optical elements and
support the steady value of optical energy circulating inside the cavities. It
can be injected into the scheme through the slightly transparent auxiliary
mirror {\sf D}.

Such system set of eigenfrequencies represents a series of doublets, with frequencies in
each doublet separated by the beating frequency
\begin{equation}
  \Omega_B = \frac{cT_{\sf C}}{L}
\end{equation}
(notations used in this paper are gathered in Table\,\ref{tab:notations}). If
the upper frequency mode of some of the doublets is pumped then optical field
acts as two rigid springs one of which is located between the mirrors ${\sf
E}_1$ and {\sf C} and the second one (L-shaped) --- between the mirrors ${\sf
E}_2$ and {\sf C}. This is the same optical rigidity that can exist in a
single cavity \cite{99a1BrKh, 01a1BrKhVo, 03a1eBiSa} and in the
signal-recycled topology of laser interferometric gravitational-wave
detectors \cite{Buonanno2002}.

Due to these springs displacement of the end mirrors ${\sf E}_{1,2}$ caused
by the gravitational wave produces displacement of the local mirror {\sf C}.
The local mirror should have an attached measurement device ({\it local meter}) which monitors its position relative to some reference mass
placed outside the optical field.

In the article \cite{02a1Kh} an improved version of the {\it optical bars}
scheme was proposed. It differs from the original ``optical bars'' scheme by
two additional mirrors ${\sf I}_1$ and ${\sf I}_2$ (see
Fig.\,\ref{fig:optbar}, right) which turn the antenna arms into two
Fabry-Perot cavities, similar to the standard Fabry-Perot---Michelson topology
of the contemporary gravitational-wave antennae. In this topology,
\begin{equation}
  \Omega_B \approx \gamma\frac{T_{\sf C}}{R_{\sf C}} \,.
\end{equation}
This scheme was called {\it optical lever} because it can provide the gain in
signal displacement of the local mirror similar to the gain which can be
obtained using ordinary mechanical lever with unequal arms. The value of this
gain is equal to
\begin{equation}
  \digamma \approx \frac{c}{\gamma L} = \frac{2}{\pi}\,{\cal F} \,.
\end{equation}
It was shown in the article \cite{02a1Kh} that in all other aspects the {\it
optical lever} scheme is identical to the {\it optical bars} one, but in the
former one the local mirror {\sf C} mass have to be $\digamma^2$ times
smaller. Due to this scaling of mass the gain in signal displacement
by itself does not allow to overcome the SQL, because the SQL value increases
exactly in the same proportion. But it allows to use less sensitive local
position meter (thus decreasing substantially required optical power in it)
and increases the signal-to-noise ratio for miscellaneous noises of
non-quantum origin.

In the article \cite{03a1Kh} prospects of use of QND local meter (mentioned
first briefly in the article \cite{98a1BrGoKh}) was analyzed. It was shown
that QND local meter allows to decrease significantly the optical power
circulating in the main cavity, while providing sensitivity several times
better than the Standard Quantum Limit.

The main goal of the current paper is further development of the {\it optical
lever} topology towards practical design of the intracavity 
gravitational-wave detector. In particular, we consider its integration with
the local meter based on the Discrete Sampling Variation Measurement
(DSVM) procedure \cite{00a1DaKhVy}.

This paper is organized as the following. 

In the Sec.\,\ref{sec:intraextra} we discuss the semi-technological
limitations mentioned above and estimate sensitivities which can be provided
by extracavity and intracavity topologies.

In the Sec.\,\ref{sec:topology} modified topology of the {\it optical lever}
scheme which can be considered as more ``practical'' one is proposed. It
differs from the  previous one by its symmetry, which allows to suppress
influence of the input light amplitude fluctuation. In addition, it provides
the means to inject optical pumping inside the scheme without increase of the
signal mode coupling with the external world ({\it i.e.} without increase of
the optical losses).

In the Sec.\,\ref{sec:optlosses} scheme potential sensitivity, {\it 
i.e.} the sensitivity limitation imposed by the optical losses, is analyzed.
In the previous papers \cite{98a1BrGoKh, 03a1Kh} this limitation was estimated
only for the simplified model based on two harmonic oscillators. Now we 
calculate it accurately. 

In the Sec.\,\ref{sec:dsvm} possible implementation of the local meter, which
is evidently the key element of the intracavity topologies, is analyzed in
detail. We consider in this section the combination of the optical lever
topology with the DSVM scheme \cite{00a1DaKhVy} and calculate its
sensitivity.

\begin{table*}[t]
  \begin{tabular}{|c|c|c|}
    \hline
		  Quantity & Value & Description \\
		\hline	
		  $A$                 & $3\times\sqrt{10^{-5}}$ & 			   
        Arm cavities amplitude loss per bounce \\ 		  
      $A_{\rm local}$     & $5\times\sqrt{10^{-6}}$ &  			   
       Local meter cavity amplitude loss per bounce \\ 		
      $c$	                &                        & Speed of light \\
			${\cal F}$          &                        & Arm cavities finesse \\
			$\digamma\approx\dfrac{2}{\pi}\,{\cal F}$ & & Signal displacement gain \\
			$\hbar$	            &                        & Plank's constant    \\
			$L$                 & 4\,km                   & Arm cavities length \\
			$l$			            &                        & Local meter cavity length\\
			$M_{\sf C}$         &                        & Central mirror mass \\
			$M_{\sf E}$         &                        & End mirrors mass    \\
			$M_{\sf I}$         &                        & Input mirrors mass  \\
			$M=\dfrac{2M_{\sf E}M_{\sf I}}{M_{\sf E}+M_{\sf I}}$ & 40\,kg &     \\
			$\mu = M/\digamma^2$ &                       &                     \\
			$m_+ = M_{\sf C}+\mu$ &             & Equivalent sum mass of the system \\
			$m_* = \dfrac{\mu M_{\rm C}}{\mu+M_{\sf C}}$ & &
			  Equivalent reduced mass of the system\\ 
			$R_{\sf C},\ T_{\sf C}$&                        & 
			  Central input mirror amplitude reflectivity and transmittance \\
			$T_{\rm local}$     &                        & 
			  Local meter input mirror amplitude transmittance \\
			$W$                 & & Optical power circulating in the arm cavities \\
			$w$			            & & Optical power circulating in local meter cavity \\
      $\gamma$            &                     & Arm cavities half-bandwidth \\
			$\gamma_{\rm loss}=\dfrac{cA^2}{4L}$ & $0.6\,{\rm s}^{-1}$ &
			  Part of $\gamma$ caused by the optical losses \\
      $\omega_o$          & $1.8\times 10^{15}\,{\rm s}^{-1}$& 
			  Optical pumping frequency \\
			$\Omega$	          & $2\pi\times 100\,{\rm s}^{-1}$ &
			  Signal (side-band) frequency \\
			$\Omega_B$          &                        & Beating frequency \\
			$\Omega_0$          &                  & Mechanical resonance frequency \\
			$\tau$			        &                        & DSVM sampling time \\
		\hline	
  \end{tabular}
	\caption{Main notations used in this paper.}\label{tab:notations}
\end{table*}

\section{Intracavity vs. extracavity topologies}\label{sec:intraextra}

\subsection{Optical power}

It is well known that in order to detect tiny gravitational-wave signal huge
amount of optical quanta is required. Usual explanation of this 
requirement is the following. In the interferometric gravitational-wave 
detectors the phase of the optical field is monitored. Precision of this 
measurement is limited by the phase quantum fluctuations ({\it
i.e.} the shot noise) which spectral density is inversely  proportional to
the mean optical power.

In the QND modifications of the standard topology, for example, variational 
input/output schemes \cite{03a1BrGoKhMaThVy, Harms2003, Buonanno2004}, not 
phase but some combination of the phase and amplitude quadratures of the
optical field is monitored. In this case more general explanation
\cite{00p1BrGoKhTh} based on the Heisenberg uncertainty relation can be 
provided. 

Really, in order to detect displacement $\sim Lh$ of the end
mirrors created by the gravitational wave it is necessary to provide
perturbation of its momentum $\Delta p \ge \hbar/2\delta x$. The only source
of this perturbation in the interferometric gravitational-wave antennae is the
uncertainty of the optical pumping energy: $\Delta p \propto \Delta{\cal E} =
\mean{{\cal E}}/\zeta^2$, where $\zeta=e^{-R}$ is the squeeze factor and
$\mean{\cal E}$ is the mean energy. Therefore, the smaller $\delta x$ have to
be  detected, the higher energy is required.

In spectral representation uncertainty relation for the interferometric
gravitational-wave detectors can be presented as the following \footnote{In 
this article, we use ``two-sided'' spectral densities which two times smaller
than ``one-sided'' ones and provide a bit more consistent formulae.}:
\begin{equation}\label{Sh_SF}
  \frac{L^2S_h}{4}\times S_{\rm B.A.} = \frac{\hbar^2}{4} \,,
\end{equation} 
where $S_h$ is the spectral density of the measurement noise, normalized as
fluctuation metrics variation, and $S_{\rm B.A.}$ is the spectral density of
the fluctuation radiation pressure differential force acting on each of the
test mirrors. 

It is evident that for all extracavity topologies $S_{\rm B.A.} \propto
W/\zeta^2$. Exact form of this spectral density depends on the specific
topology. For the ordinary Initial LIGO topology 
\begin{equation}\label{S_BA}
  S_{\rm B.A.} = \frac{8\hbar\omega_pW}{\zeta^2cL}\,
	  \frac{\gamma}{\gamma^2+\Omega^2} \,, 
\end{equation}
and therefore
\begin{equation}\label{EQL}
  S_h = \frac{\zeta^2\hbar c}{8L\omega_pW}\,
	  \frac{\gamma^2+\Omega^2}{\gamma}\,.
\end{equation}
It is convenient to compare this spectral density with the one corresponding 
to the Standard Quantum Limit:
\begin{equation}\label{xi2}   
  \xi_{\rm extra}^2 \equiv \frac{S_h}{S_h^{\rm SQL}} 
  = \frac{\zeta^2}{2}\,\frac{W_{\rm SQL}}{W}\, 		  
      \frac{\gamma^2+\Omega^2}{2\gamma\Omega} \,, 
\end{equation} 
where
\begin{equation}\label{S_SQL}
  S_h^{\rm SQL} = \frac{4\hbar}{M\Omega^2L^2}
\end{equation}
(see \cite{03a1BrGoKhMaThVy}), and
\begin{equation}\label{W_SQL}
  W_{\rm SQL} = \frac{McL\Omega^3}{8\omega_o}
\end{equation}
is the circulating optical power in the SQL-limited detector which is
necessary  to reach the SQL. Factor $1/2$ corresponds to the evident fact that
QND  techniques provide $\sqrt{2}$ times better sensitivity than SQL-limited 
detector even if $W=W_{\rm SQL}$, because they ``filter out'' back-action 
noise which for $W=W_{\rm SQL}$ corresponds to one half of the total noise.

In the {\it speed-meter} topologies \cite{00a1BrGoKhTh, Purdue2001, Purdue2002,
Chen2002, 02a2Kh, 04a1Da} $S_{\rm B.A.}$ differs only by an additional factor
$2\Omega^2/(\gamma^2+\Omega^2)$. Therefore, if $\Omega\simeq\gamma$ then
sensitivity is close to one defined by Eq.\,(\ref{EQL}). 

On the other hand, in the signal recycled ``optical springs'' topology
\cite{Buonanno2001, 01a2Kh, Buonanno2002, Buonanno2003, Harms2003,
Buonanno2004} it is possible to create high narrow peak in spectral
dependence of $S_{\rm B.A.}$:
\begin{equation}
  S_{\rm B.A.} = \frac{4\hbar\omega_pW}{\zeta^2cL}\,
	  \frac{\Delta\Omega/2}{(\Omega-\Omega_0)^2+(\Delta\Omega/2)^2} \,.
\end{equation}
The peak width $\Delta\Omega$ and the mean frequency $\Omega_0$ depend on the 
signal recycling mirror cavity parameters. Therefore, in this case it is
possible to obtain sensitivity much better than the SQL without increase of
optical power, but only in narrow spectral band $\Delta\Omega\ll\Omega_0$:
\begin{align}
  \xi_{\rm extra}^2 &= \frac{\zeta^2}{2}\,\frac{W_{\rm SQL}}{W}\,
	  \frac{(\Omega-\Omega_0)^2+(\Delta\Omega/2)^2}{\Omega_0\Delta\Omega/2}\,,&
	\xi_{\rm extra}^2\Bigr|_{\Omega=\Omega_0} 
	  &=\frac{\zeta^2}{2}\,\frac{W_{\rm SQL}}{W}\,
     	  \frac{\Delta\Omega/2}{\Omega_0}	\,.
\end{align} 
Below we limit ourselves to the wide-band case (\ref{xi2}) only.

\subsection{Optical losses}

It follows from Eq\,(\ref{xi2}) that the best sensitivity can be achieved if 
$\gamma\simeq\Omega$, and at this point 
\begin{equation}\label{xi2_best}
  \xi_{\rm meter}^2 = \frac{\zeta^2}{2}\,\frac{W_{\rm SQL}}{W}\,.
\end{equation}
Unfortunately, situation is possible where this optimization can not be
provided. Really, it can be shown that internal losses in the optical elements
impose the following additional limitation on the sensitivity:
\begin{equation}\label{xi2loss_org}
  \xi_{\rm loss}^2 
	  = \sqrt{\zeta^2\frac{\gamma_{\rm loss}}{\gamma_{\rm load}}} 
    \approx \sqrt{\zeta^2\frac{\gamma_{\rm loss}}{\gamma}} \,.
\end{equation}
Suppose that $\gamma\approx\Omega$. In this case sensitivity will be
limited by the following value:
\begin{equation}
  \xi_{\rm loss}^2 \approx \sqrt{\zeta^2\frac{\gamma_{\rm loss}}{\Omega}} \,.
\end{equation}
For the Advanced LIGO values of parameters (see Table\,\ref{tab:notations}),
\begin{equation}
  \gamma_{\rm loss} = \frac{cA^2}{4L} \simeq 0.6\,{\rm s}^{-1} \,,
\end{equation}
and 
\begin{equation}
  \xi_{\rm loss} \approx 0.2\sqrt{\zeta} \,.
\end{equation}
In order to obtain smaller $\xi_{\rm loss}$ it is necessary to increase
$\gamma$ thus increasing $\xi_{\rm meter}$. It is evident that the optimal 
value of $\gamma$ exists which provides minimum to the sum noise spectral 
density:
\begin{equation}\label{xi2sum_raw}
  \xi_{\rm sum}^2 = \xi_{\rm extra}^2 + \xi_{\rm loss}^2
	  \approx\frac{\zeta^2}{2}\,\frac{W_{\rm SQL}}{W}\,\frac{\gamma}{2\Omega} 
	  + \sqrt{\zeta^2\frac{\gamma_{\rm loss}}{\gamma}} 
\end{equation}
(it is supposed here for simplicity that $\gamma\gg\gamma_{\rm loss}$, 
$\gamma\gg\Omega$). The minimum is reached when 
\begin{equation}
  \gamma = \left(\frac{4\gamma_{\rm loss}\Omega^2}{\zeta^2}\,
		\frac{W^2}{W_{\rm SQL}^2}\right)^{1/3} \,,
\end{equation}
and is equal to:
\begin{equation}\label{xi2sum}
  \xi_{\rm sum}^2 = \frac{3}{2}\left(\zeta^4\frac{\gamma_{\rm loss}}{2\Omega}\, 	  
	  \frac{W_{\rm SQL}}{W}\right)^{1/3} \,. 
\end{equation}
For the values of $\gamma_{\rm loss}$ and $\Omega$ mentioned above, we 
obtain, that
\begin{equation}
  \xi_{\rm sum} \approx 0.34\times\zeta^{2/3}\times
	  \left(\frac{W_{\rm SQL}}{W}\right)^{1/6} \,.
\end{equation}
Note very weak dependence on pumping power.

\begin{figure*}[t]
  \psfrag{0}[lb][lb]{(a)}
  \psfrag{10}[lb][lb]{(b)}
  \psfrag{20}[lb][lb]{(c)}
  \psfrag{P}[lb][lb]{(d)}
  \psfrag{V}[lt][lb]{(e)}
  \psfrag{D}[lb][lb]{(f)} \psfrag{w}[ct][ct]{$W/W_{\rm SQL}$}
  \psfrag{xi}[cb][lb]{$\xi=\sqrt{\dfrac{S_h}{S_h^{\rm SQL}}}$}
  \includegraphics[width=0.6\textwidth]{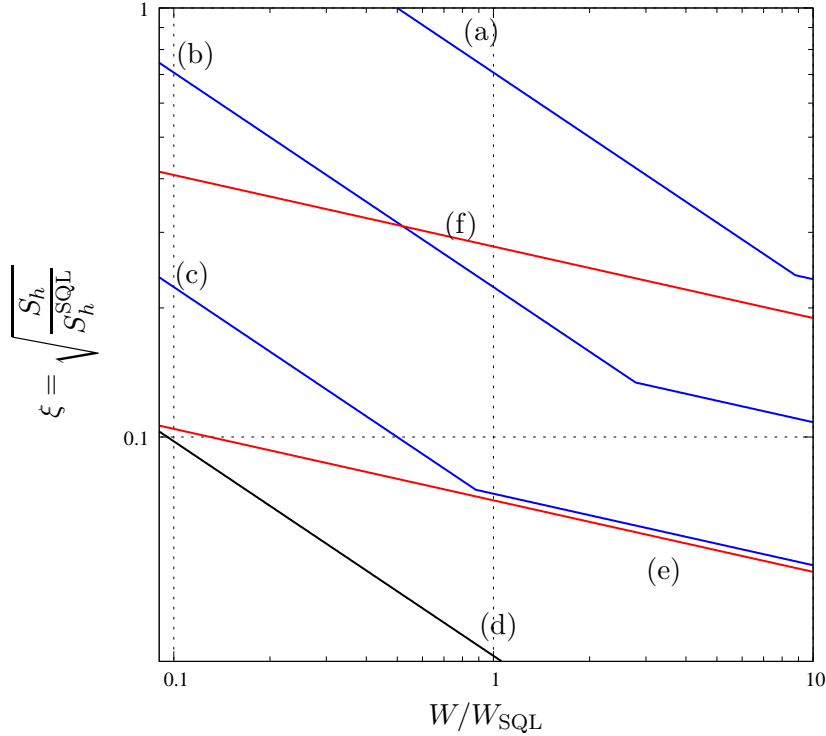}
  \caption{(a): Sensitivity which can be obtained in the standard extracavity
	topology for the coherent pumping; (b): the same for the $10\,{\rm dB}$
  squeezed pumping; (c): $20\,{\rm dB}$ squeezed pumping. The left more steep
  parts of curves (a)-(c) correspond to Eq.\,(\ref{xi2}) with $\gamma=\Omega$,
  the right more flat ones --- to Eq.\,(\ref{xi2sum}). (d): Potential
  sensitivity of the intracavity optical lever topology; (e) sensitivity of
  the optical lever scheme with the spectral variation measurement based local meter; (f)
  sensitivity of the optical lever scheme with the DSVM-based local meter.}  
	\label{fig:sensitivity}.
\end{figure*}

The sensitivity estimates based on Eqs.\,(\ref{xi2}),\,(\ref{xi2sum}) are 
plotted in Fig.\,\ref{fig:sensitivity} as functions of optical power, see
curves (a),(b),(c). 

\section{Practical version of the \textit{optical lever} intracavity
topology}

\subsection{Discussion of the topology}\label{sec:topology}

\begin{figure*}[t]
  \psfrag{E1}[cb][lb]{${\sf E}_1$}
  \psfrag{E2}[rc][lb]{${\sf E}_2$}
  \psfrag{I1}[cb][lb]{${\sf I}_1$}
  \psfrag{I2}[rc][lb]{${\sf I}_2$}
  \psfrag{C}[lb][lb]{{\sf C}}
  \psfrag{D1}[lb][lb]{${\sf D}_1$}
  \psfrag{D2}[lb][lb]{${\sf D}_2$}
  \psfrag{P1}[ct][lt]{${\sf P}_1$}
  \psfrag{P2}[ct][lt]{${\sf P}_2$}
  \psfrag{S}[lb][lb]{{\sf S}}
  \psfrag{BS}[rt][lb]{{\sf BS}}
  \includegraphics[width=0.6\textwidth]{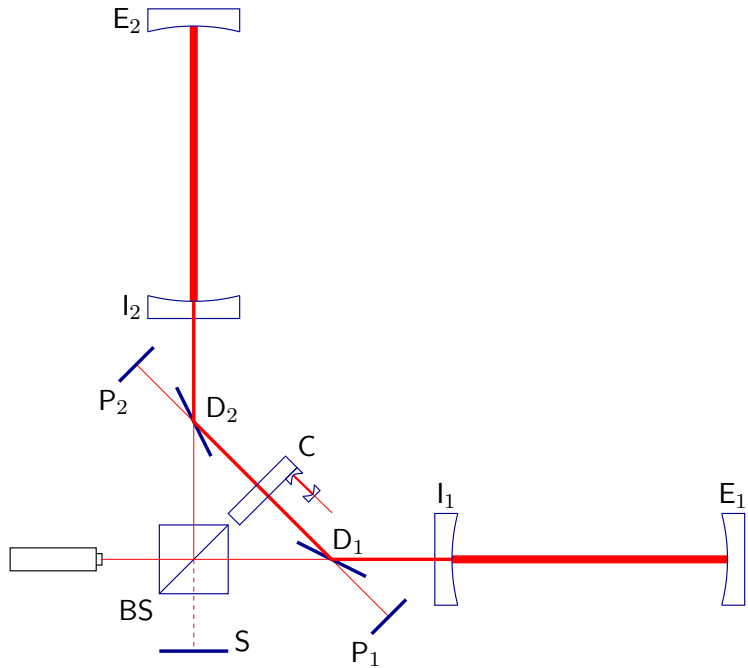}
  \caption{Practical design of the {\it optical lever} intracavity
  scheme}\label{fig:topology}
\end{figure*}

The scheme which is analyzed in this paper is presented in Fig.\,\ref{fig:topology}. Consider step by step the additional  optical
elements of this scheme.

\paragraph{Symmetrization of the topology.}

The evident disadvantage of simple schemes shown in Fig.\,\ref{fig:optbar} is
their non-symmetry: pumping power enters  the ``north'' (vertical on the
picture) arm first and only then, through the coupling mirror {\sf  C}, the
``east'' one. Due to this non-symmetry the input optical field amplitude
fluctuations will create differential pondermotive  force acting on the
central mirror and imitating gravitational-wave signal. In order to eliminate
this effect, symmetric power injection scheme shown in
Fig.\,\ref{fig:topology} have to be used. It consists of the beamsplitter {\sf
BS} which splits the input beam into two and two power injection mirrors ${\sf
D}_1$ and ${\sf D}_2$ placed symmetrically on both sides of the central 
mirror {\sf C}. 

\paragraph{Power recycling mirrors.}

It can be shown that without power recycling mirrors ${\sf P}_1$, ${\sf
P}_2$ one quarter of input power is reflected from the mirrors ${\sf
D}_1$ and ${\sf D}_2$ back to the laser, another quarter is reflected to
the side direction, and only one half enters the scheme. The mirrors
${\sf P}_1$, ${\sf P}_2$ cancel both reflected beams and increase twice the
circulating power inside the scheme (for the same value of input power). 

\paragraph{Signal recycling mirror}
 
It can be shown also that if the mirrors ${\sf D}_{1,2}$ transmittances are
tuned in optimal way to provide maximal optical power in the scheme [see 
Eq.\,\ref{T_D_opt}] then these transmittances will create an additional
``hole'' which will increase two-fold total optical losses in the scheme. 

This ``hole'' can be closed without affecting optimal coupling condition 
using symmetry of the scheme. Indeed, similar to traditional 
interferometric gravitational-wave detectors topology, the mean value of 
optical power inside the scheme depends on the bandwidth of the symmetric 
optical mode which is coupled with ``western'' port of the beamsplitter, 
and the detector sensitivity depends on the bandwidth of anti-symmetric 
mode which is coupled with ``south'' port of the beamsplitter. The only 
difference is that in traditional topology the anti-symmetric mode
bandwidth have to be close to the signal frequency $\Omega$ to provide optimal coupling  with photodetector, while in the intracavity topology it
have to be as small as possible. Therefore, high-reflectivity signal
recycling mirror {\sf S}  have to be placed in the ``south'' port as shown in
Fig.\,\ref{fig:topology}.

\subsection{Sensitivity limitation due to optical losses}
\label{sec:optlosses}

The topology described in the previous subsection is analyzed in the Appendix 
\ref{app:topology}. In particular, the sensitivity limitation imposed by optical losses is calculated. Spectral density of
the corresponding equivalent noise (normalized as fluctuation metrics
variation) is equal to: 
\begin{equation}\label{opt_loss_simple}
  S_h^{\rm loss}
	\approx\frac{\hbar c\gamma_{\rm loss}}{2\omega_oWL}
	    \left(1+\frac{\Omega^2}{\Omega_B^2}\right) \,. 
\end{equation}
(slightly simplified form is presented here, which takes into account that 
$\Omega_B\ge\Omega\gg\gamma_{\rm loss}$; for the exact form, see
Eq.\,(\ref{opt_loss}).).

Compare this spectral density with the one corresponding 
to the Standard Quantum Limit [see Eqs.\,(\ref{S_SQL}),\,(\ref{W_SQL})]:
\begin{equation}\label{xi2loss}
  \xi^2_{\rm loss}\equiv \frac{S_{h\,\rm loss}}{S_h^{\rm SQL}}
	= \frac{\gamma_{\rm loss}}{\Omega}\,\frac{W_{\rm SQL}}{W} 
	    \left(1+\frac{\Omega^2}{\Omega_B^2}\right) \,.
\end{equation}
It was noted in the article \cite{03a1Kh}, that due to the fact that factor
$\gamma_{\rm loss}/\Omega$ can be as small as $\sim 10^{-3}$, the value
$\xi_{\rm loss}\ll 1$ can be obtained even with $W\ll W_{\rm SQL}$.

Estimate of $\xi_{\rm loss}$ as a function of $W/W_{\rm SQL}$ (the potential
sensitivity) is plotted in  Fig.\,\ref{fig:sensitivity}, see curve (d).

\section{Local meter}\label{sec:dsvm}

\subsection{Options for the local meter}  Taking into account the gain
$\digamma\sim 10\div100$ in the local mirror
mechanical displacement, sensitivity of the local meter have to be several
times better  than SQL for the mass $\mu=M/\digamma^2$:
\begin{equation}
  \sqrt{\frac{\hbar}{\mu\Omega^2}} = \digamma\sqrt{\frac{\hbar}{M\Omega^2}}
	  \sim (10\div 100)\times2.5\times 10^{-19}\,{\rm m\times s^{-1/2}}\,.
\end{equation}
Several types of devices have been proposed which can, in principle,
provide  this sensitivity, in particular: squeezed-based schemes used in
solid-state  gravitational-wave antennae; microwave speed-meter
\cite{00a1BrGoKhTh}; small-scale optical speed-meter \cite{Chen2002}; spectral 
variation measurement-based schemes ({\it a.k.a.} schemes with modified 
input-output optics) \cite{02a1KiLeMaThVy, Buonanno2004}; and the Discrete
Sampling Variation Measurement (DSVM) based optical position meter
\cite{00a1DaKhVy}.

The first two types require cryogenic equipment. In addition, estimates 
made in the article \cite{00a1BrGoKhTh} show that due to the internal 
losses the microwave speed-meter can provide sensitivity only slightly better 
than SQL. 

The Sagnac-based optical speed-meter (as well as other ``practical'' speed-meter
schemes) requires that its optical storage time has to be larger 
than $\Omega^{-1}\sim 10^{-3}\,{\rm s}$. Simple estimates show that due to 
this limitation the interferometer size can not be smaller than $\sim 100\,{\rm 
m}$, {\it i.e.} an additional setup comparable with a full scale 
gravitational-wave detector, such as GEO-600, is necessary.

In spectral variation measurement based (variational input/output) schemes a
short (desktop-scale) main cavity can be used. However, they require an
additional cavity with bandwidth  comparable with the signal frequency and
thus with hundreds meters length. 

We consider here two variants of the local meter: the spectral variation
measurement based and DSVM-based schemes. The former one, while can not be 
considered as a candidate for a practical implementation, allows, in 
principle, to obtain the best sensitivity and thus can be considered as an 
ideal ``asymptotic case'' for all other schemes. The DSVM-based local meter
can be considered as a realistic scheme but its sensitivity, unfortunately,
is by far not so good just due to a couple of peculiar numeric factors specific 
for this scheme.

\begin{figure}
  \psfrag{C}[cb][lb]{{\sf C}}
  \includegraphics[width=0.6\textwidth]{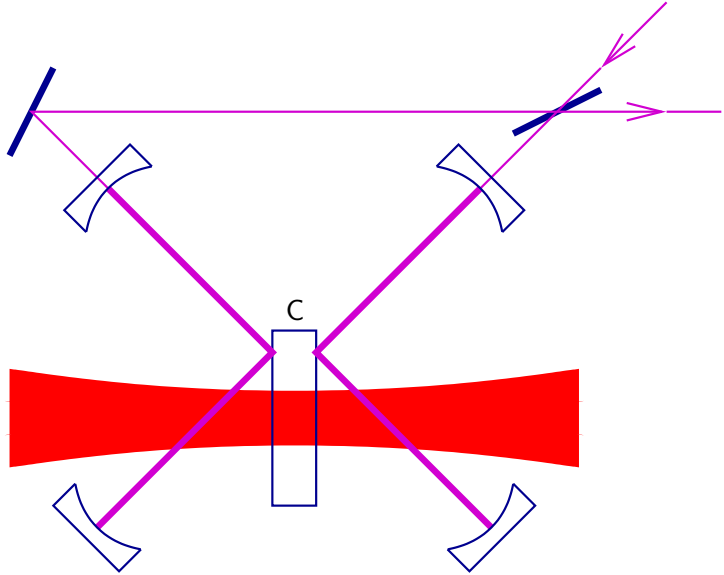}
  \caption{Possible design of local meter}\label{fig:local}
\end{figure}

Both these schemes use Fabry-Perot cavity-based position meter with a homodyne
detector. The evident technical challenge in this case is how to attach this
meter to the small  (with the mass of about 1 gram) local mirror which is
also the part of the  main large-scale optical setup. Possible solution which
is based on the  scheme proposed in the paper \cite{Corbitt2004} is shown in 
Fig.\,\ref{fig:local}.

\subsection{Ideal variation measurement}

Suppose that the local oscillator phase $\phi_{\rm LO}$ of the homodyne
detector mentioned above can depend on the observation frequency $\Omega$ in
an arbitrary way. It was shown in the article \cite{02a1KiLeMaThVy} that by
special tuning of the function $\phi_{\rm LO}(\Omega)$ it is possible to
eliminate the back-action noise from the output signal and thus to overcome
the SQL.

Spectral density of this scheme measurement noise is calculated in Appendix 
\ref{app:KLMTV}, see Eq.\,(\ref{S_var}). It follows from this equation that 
the sensitivity limitation imposed by the meter can be presented as follows:
\begin{equation}\label{xi2_var_raw}
  \xi_{\rm meter}^2 \equiv \frac{S_h^{\rm meter}}{S_h^{\rm SQL}}
	= \frac{{\cal I}}{2}\,\frac{m_+^2}{\mu M_{\rm C}}\,\frac{w_{\rm SQL}}{w} \,,
\end{equation}
where
\begin{equation}
  {\cal I} = \frac{\left[\Omega^4-\Omega^2\Omega_B^2+\Omega_0^2\Omega_B^2\right]^2}
	  {\Omega_0^4\Omega_B^4}\,,
\end{equation}
and
\begin{equation}
  w_{\rm SQL} = \frac{M_{\sf C}c^2T_{\rm local}^2\Omega^2}{32\omega_o}
\end{equation}
is circulating power in an ordinary (SQL-limited) Fabry-Perot cavity-based 
position meter which is necessary to reach the SQL for the test mass $M_{\bf 
C}$.

Factor ${\cal I}$ has rather sophisticated spectral dependence. It is evident, 
however, that the best sensitivity area corresponds to values 
$\Omega\sim\Omega_0\sim\Omega_B$, and the noise spectral density increases as 
$\Omega^4$ if $\Omega\gg\Omega_B\sim\Omega_0$. 

We consider here simple particular case when
\begin{equation}\label{small_I}
  \Omega \le \Omega_B = 2\sqrt{2}\Omega_0 \,.
\end{equation}
(for more general optimization, see Appendix C of the article \cite{03a1Kh}).
In this case ${\cal I}\le 1$. On the other hand, condition (\ref{small_I})
together with Eq.\,(\ref{Omega0}) lead to the following limitation  on the
pumping power $W$:
\begin{equation}\label{W_var_raw}
  W\ge\frac{\digamma^2m_*cL\Omega^3}{64\omega_o}
	  =\frac{1}{8}\frac{m_*}{\mu}\,W_{\rm SQL} \,.
\end{equation}
It was pointed in the article \cite{03a1Kh} that it is possible to reduce 
pumping power by using small local mirror with mass $M_{\sf C}\ll\mu$. In 
this case, 
\begin{align}
  m_+ &\approx \mu\,, & m_* &\approx M_{\sf C} \,,
\end{align}
and Eqs.\,(\ref{xi2_var_raw}),\,(\ref{W_var_raw}) can be simplified:
\begin{subequations}\label{xiW_var}
  \begin{gather}
    \xi^2_{\rm meter} = \frac{{\cal I}}{2\digamma^2}\,\frac{M}{M_{\rm C}}\,
      \frac{w_{\rm SQL}}{w}\,,\\
    W \ge \frac{\digamma^2}{8\,}\frac{M_{\sf C}}{M}\,W_{\rm SQL} \,,
  \end{gather}
\end{subequations}

The meaning of these equations is evident. The larger is $\digamma$, the
better is sensitivity because the local mirror signal displacement is
proportional to $\digamma$. On the other hand,  the larger is $\digamma$, the
larger have to be circulating power in the arm  cavities to keep optical
springs sufficiently stiff. Excluding factor $\digamma$ equations 
(\ref{xiW_var}) can be combined into the following one:
\begin{equation}\label{xi2_var}
  \xi^2_{\rm meter} = \frac{1}{16}\,\frac{W_{\rm SQL}}{W}\,
    \frac{w_{\rm SQL}}{w}\,.
\end{equation}

In Eq.\,(\ref{xi2_var}) optical losses in the local meter cavity have not been 
taken into account. These losses impose an additional sensitivity limitation, 
which have the same form as condition (\ref{xi2loss_org}):
\begin{equation}\label{xi2local_loss}
  \xi^2_{\rm loss}	= \sqrt{\frac{A_{\rm local}^2}{T_{\rm local}^2}}\,.
\end{equation}
The smaller is $T_{\rm local}$, the smaller is $\xi_{\rm meter}$, but the
larger is $\xi_{\rm loss}$. Therefore, an optimal value of $T_{\rm  local}$
exists where the sum 
\begin{equation}
  \xi^2_{\rm meter\,loss} = \xi^2_{\rm meter}	+ \xi^2_{\rm loss}
\end{equation}
is minimum. It is easy to show that at this point,
\begin{equation}\label{xi2var_loss}
  \xi^2_{\rm meter\,loss} = \frac{\xi_0^2}{2}
	  \left(\frac{W_{\rm SQL}}{W}\right)^{1/3} \,,
\end{equation}
where
\begin{equation}	
  \xi_0^2 = \frac{3}{2}\left(
	  \frac{M_{\sf C}c^2A_{\rm local}^2\Omega^2}{32\omega_ow}
   \right)^{1/3} \,.
\end{equation}

For numeric estimates,we will use the same values as proposed for the
pondermotive squeezing experiment in \cite{Corbitt2004}, see
Table\,.\ref{tab:notations}. For these values $\xi_0\approx 0.1$ thus 
allowing to obtain for the optical power $W\lesssim W_{\rm SQL}$ the value of
$\xi_{\rm meter\,loss}$ which is also close to 0.1. Graphics of $\xi_{\rm
meter\,loss}$ as a function of $W$ is plotted in Fig.\,\ref{fig:sensitivity}, 
see curve (e).

\subsection{DSVM-based local meter}  

The scheme of the DSVM-based local meter, similar to the previous one, 
consists of a Fabry-Perot cavity-based position meter with a homodyne
detector. However, instead of the frequency-dependent local oscillator phase
time-dependent one is used in order to exclude the back-action noise.

This method is based on variation measurement technique proposed in
\cite{95a1VyZu} and analyzed in \cite{96a2eVyMa, 98a1Vy}. Severe disadvantage
of this original form of variation  measurement is the necessity to know the
shape and arrival time of the signal being detected. DSVM procedure suggests
the way to overcome this disadvantage by approximating the real signal with the
sequence of rectangular pulses which amplitude is the mean value of the signal
over the pulse duration $\tau\le\pi/\Omega_{\rm max}$, where $\Omega_{\rm
max}$ is the upper frequency  of the signal.

Sensitivity of the DSVM-based local meter is calculated in Appendix
\ref{app:dsvm}. It is shown that if this meter is used then 
\begin{equation}\label{xi_DSVM}
  \xi_{\rm DSVM}^2 \equiv \frac{S_h^{\rm meter}}{S_h^{\rm SQL}}
  = \frac{720}{\pi^4{\cal G}(\Omega_B\tau,\Omega_0\tau)}\,
  	  \frac{m_+^2}{\mu M_{\sf C}}\,\frac{w_{\rm SQL}}{w} 
\end{equation} 
(it is supposed here that $\Omega=\Omega_{\rm max}$).

\begin{figure*}[t]      
  \psfrag{xi}[cc][cc]{${\cal G}(\Omega_B,\,\Omega_0)$}     
	\psfrag{W}[cc][cc]{$\Omega_B\tau$}    
  \psfrag{L}[cc][cc]{$\Omega_0\tau$}      
  \includegraphics[width=0.7\textwidth]{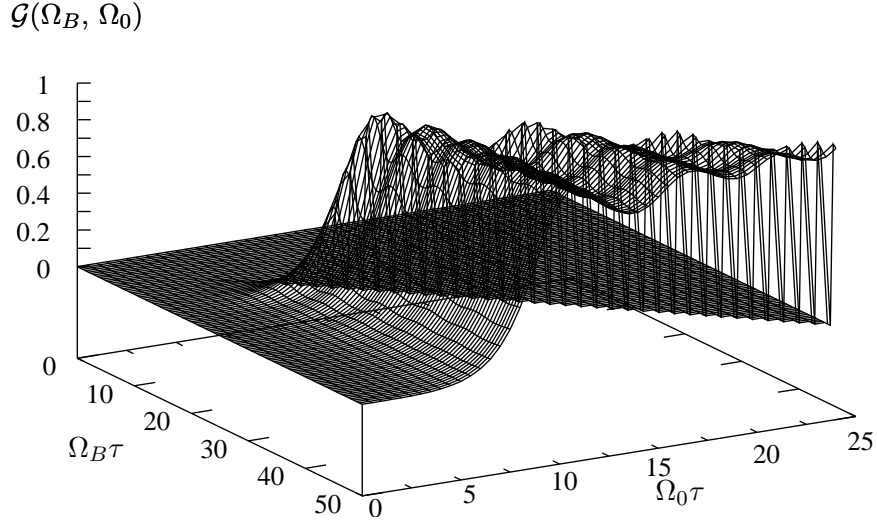}    
  \caption{3D plot of function ${\cal G}(\Omega_B\tau,\Omega_0\tau)$.}
	\label{fig:F_3D}  
\end{figure*}  

Dimensionless function ${\cal G}(\Omega_B\tau,\Omega_0\tau)$ is calculated 
numerically and its 3D-plot is presented in Fig.\ref{fig:F_3D}. Three areas
can be clearly distinguished on this plot depending on the mechanical
eigenfrequency $\Omega_0$ and beating frequency $\Omega_B$.

1. $\Omega_0>\Omega_B/2$. In this area the system is extremely unstable: its 
eigenfrequencies have imaginary parts of both signs comparable with the real
ones. We expressed symbolically this instability by setting ${\cal G}=0$ ({\it 
i.e.} $S_h^{\rm meter}\to\infty$) in this area. 

2. $\Omega_B/2>\Omega_0\gtrsim 3\Omega_{\rm max}$. In this area, ${\cal G}$ 
is close to its maximum value 1 and therefore the best sensitivity is 
provided. Condition $\Omega_0\gtrsim 3\Omega_{\rm max}$ describes 
sufficiently stiff optical springs that provide the local mirror signal 
displacement equal to the end mirrors displacement multiplied by factor 
$\digamma$. 

3. $\Omega_0\lesssim 3\Omega_{\rm max}$. In this area optical springs are 
too weak to move local mirror effectively. In this case the local mirror 
displacement is proportional to the rigidity $\Omega_0^2$ and 
the noise spectral density (\ref{S_meter}) to 
$\Omega_0^{-4}$, correspondingly. 
%The asymptotic case $\Omega_0\tau=\pi\Omega_0/\Omega_{\rm
%max}\ll 1$ is calculated in Appendix \ref{app:small_power} and indeed this 
%dependence is obtained, see Eq.\,(\ref{S_meter_small}) 

Below we consider the best sensitivity case where the condition 
\begin{equation}\label{flat_G}
  \Omega_B/2>\Omega_0\gtrsim 3\Omega_{\rm max}
\end{equation}
is fulfilled and thus
\begin{equation}\label{xi2_DSVM_raw}
  \xi^2_{\rm DSVM} \approx \frac{720}{\pi^4}\,\frac{m_+^2}{\mu M_{\sf C}}\, 	 
    \frac{w_{\rm SQL}}{w} \,.
\end{equation} 
On the other hand, condition (\ref{flat_G}) together with Eq.\,(\ref{Omega0})
lead to the limitation  on the pumping power $W$:
\begin{equation}\label{W_DSVM_raw}
  W\gtrsim 2\times3^3\times\frac{\digamma^2m_*cL\Omega^3}{8\omega_o}
	  \approx 60\frac{m_*}{\mu}\,W_{\rm SQL} \,.
\end{equation}

Note that Eqs.\,(\ref{xi2_DSVM_raw}),(\ref{W_DSVM_raw}) have exactly the 
same structure as Eqs.\,(\ref{xi2_var_raw}),(\ref{W_var_raw}) for the ideal 
meter case and differ by numerical factors only. Therefore, the next 
consideration follows the previous subsection.

We suppose again that $M_{\sf C}\ll\mu$ and thus obtain that:
\begin{subequations}
  \begin{gather}
    \xi^2_{\rm DSVM}\approx \frac{720}{\pi^4\digamma^2}\,
      \frac{M}{M_{\sf C}}\,\frac{w_{\rm SQL}}{w} \,,\\
    W \approx 60\digamma^2\frac{M_{\sf C}}{M}\,W_{\rm SQL} \,.
  \end{gather}
\end{subequations}
Combining again these two equation we obtain the following formula for
the DSVM-based local meter:
\begin{equation}\label{xi2_DSVM}
  \xi^2_{\rm DSVM} \approx \frac{720\times60}{\pi^4}\,
	  \frac{W_{\rm SQL}}{W}\,\frac{w_{\rm SQL}}{w}\,.
\end{equation}
The final step is again optimization of $T_{\rm local}$ which gives that:
\begin{equation}\label{xi2DSVM_loss}
  \xi^2_{\rm DSVM\,loss} \approx \xi_0^2
	  \left(\frac{720\times60}{\pi^4}\frac{W_{\rm SQL}}{W}\right)^{1/3} \,,
\end{equation}
Graphics of $\xi_{\rm DSVM\,loss}$ as a function of $W$ is also plotted in
Fig.\,\ref{fig:sensitivity}, see curve (f).

\section{Conclusion}

Comparing traditional extracavity topologies and  intracavity 
topologies discussed in this article, one can conclude that the possibility to obtain sensitivity 
substantially better than the Standard Quantum Limit in both cases depends in a 
crucial way on additional ``supporting'' device: squeezed state        
generator for traditional topologies, and the local meter for intracavity ones.  

In both cases the best design of the ``supporting'' device, from the
contemporary point of view, is based on small-scale Fabry-Perot cavity, with 
approximately the same reqirements to the parameters. 

However, intracavity topologies promise significantly better sensitivity,
especially for the relatively small values of pumping power: $W<W_{\rm SQL}$.
Unfortunately, none of the mechanical QND schemes known today which can be 
considered as practical ones, can fully realize this high potential
sensitivity of intracavity topologies.  From the authors point of view the search
of new methods of mechanical QND  measurements probably based on improved
DSVM scheme or which combine the local meter  with the pondermotive squeezing
techique, is necessary.

\acknowledgments

This work was supported in part by NSF and Caltech grant PHY-0353775, by
Russian Agency of Industry and Science contracts \#40.02.1.1.1.1137 and 
\#40.700.12.0086, and by Russian Foundation for Basic Research grant 
\#03-02-16975-a. 

\appendix

\section{Derivation of the mechanical equations of motion}\label{app:topology}

\begin{figure*}[t]
  \psfrag{a1}[cb][lb]{${\rm a}_1$}\psfrag{a2}[lc][lb]{${\rm a}_2$}
  \psfrag{b1}[cb][lb]{${\rm b}_1$}\psfrag{b2}[lc][lb]{${\rm b}_2$}
  \psfrag{c1}[lb][lb]{${\rm c}_1$}\psfrag{c2}[lb][lb]{${\rm c}_2$}
  \psfrag{d1}[lb][lb]{${\rm d}_1$}\psfrag{d2}[lb][lb]{${\rm d}_2$}
  \psfrag{e1}[ct][lb]{${\rm e}_1$}\psfrag{e2}[rc][lb]{${\rm e}_2$}
  \psfrag{f0}[cb][lb]{${\rm f}_0$}
  \psfrag{f1}[ct][lb]{${\rm f}_1$}\psfrag{f2}[rc][lb]{${\rm f}_2$}
  \psfrag{g1}[rt][lb]{${\rm g}_1$}\psfrag{g2}[rt][lb]{${\rm g}_2$}
  \psfrag{h1}[rt][lb]{${\rm h}_1$}\psfrag{h2}[rt][lb]{${\rm h}_2$}
  \psfrag{i1}[cb][lb]{${\rm i}_1$}\psfrag{i2}[lc][lb]{${\rm i}_2$}
  \psfrag{j1}[ct][lb]{${\rm j}_1$}\psfrag{j2}[rc][lb]{${\rm j}_2$}
  \psfrag{n1}[cb][lb]{${\rm n}_1$}\psfrag{n2}[lc][lb]{${\rm n}_2$}
  \psfrag{E1}[cb][lb]{${\sf E}_1$}
  \psfrag{E2}[rc][lb]{${\sf E}_2$}
  \psfrag{I1}[cb][lb]{${\sf I}_1$}
  \psfrag{I2}[rc][lb]{${\sf I}_2$}
  \psfrag{C}[rt][lb]{{\sf C}}
  \psfrag{A1}[rc][lb]{${\sf D}_1$}
  \psfrag{A2}[ct][lb]{${\sf D}_2$}
  \psfrag{P1}[ct][lt]{${\sf P}_1$}
  \psfrag{P2}[ct][lt]{${\sf P}_2$}
  \psfrag{S}[lb][lb]{{\sf S}}
  \psfrag{BS}[rt][lb]{{\sf BS}}
  \psfrag{a}[lc][lb]{$\beta''$}
  \psfrag{ia}[cb][lb]{$\beta''+\dfrac{\pi}{2}$}
  \psfrag{b}[lb][lb]{$\beta'$}
  \psfrag{Q11}[cb][lb]{$\theta'$}\psfrag{Q12}[lc][lb]{$\theta'$}
  \psfrag{Q2}[lb][lb]{$\theta''$}
  \psfrag{xi1}[lc][lb]{$x_{{\sf I}1}$}\psfrag{xi2}[rc][lb]{$x_{{\sf I}2}$}
  \psfrag{xe1}[lc][lb]{$x_{{\sf E}1}$}\psfrag{xe2}[rc][lb]{$x_{{\sf E}2}$}
  \psfrag{y}[lb][lb]{$y$}
  \includegraphics[width=\textwidth]{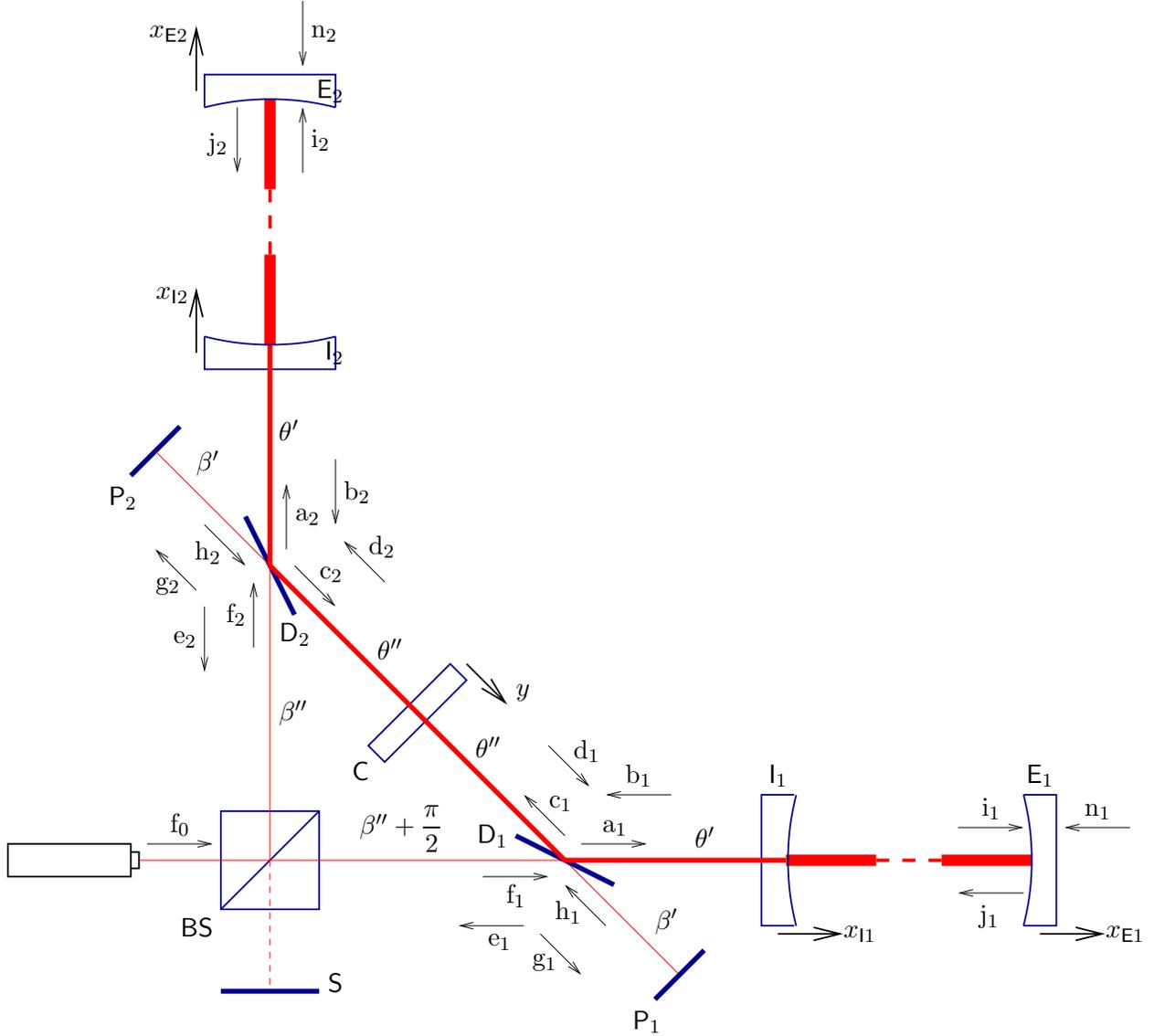}
  \caption{The scheme}\label{fig:PL_scheme}
\end{figure*}

\subsection{Notations and approximations}

\begin{table*}
  \begin{tabular}{|c|c|}
    \hline
	  Quantity & Description \\
  	\hline	
    ${\rm a_{1,2} - j_{1,2}}$ & Field amplitudes, 
	    see Fig.\,\ref{fig:PL_scheme} (roman letters are used) \\
	  ${\rm A_{1,2} - J_{\rm 1,2}}$	& Corresponding mean (classical) values 
		  (capital roman letters are used) \\
		${\rm n_{1,2}}$	& Noises created by optical losses \\
    $R_{\sf E}$, $R_{\sf I}$, {\it etc} & Amplitude reflectivities of the 
		  mirrors \\
    $T_{\sf E}$, $T_{\sf I}$, {\it etc} & Amplitude transmittances of the 
      mirrors \\
    $l_{\sf C-D}$, $l_{\sf D-I}$, {\it etc} & Optical distances between the 
		  corresponding optical elements \\
		\hline	
  \end{tabular}
  \begin{align*}%\label{thetabeta}
    \theta' &= \frac{\omega_o l_{\sf D-I}}{c} \hspace{-0.5em}\mod 2\pi \,, \\    
    \theta'' &= \frac{\omega_o l_{\sf C-D}}{c} \hspace{-0.5em}\mod 2\pi \,, \\ 
    \theta &= \theta'+\theta'' \,, \\
    \beta' &= \frac{\omega_o l_{\sf D-P}}{c} \hspace{-0.5em}\mod 2\pi \,, \\   
    \beta'' &= \left(\frac{\omega_o l_{\sf BS-P_1}}{c}-\frac{\pi}{2}\right) 
      \hspace{-0.5em}\mod 2\pi 
      = \frac{\omega_o l_{\sf BS-P_2}}{c} \hspace{-0.5em}\mod 2\pi \,.
  \end{align*}
	\caption{Some additional notations not listed in Table\,\ref{tab:notations}}
	  \label{tab:notation2}.
\end{table*}

Additional notations used in this Appendix and not listed in 
Table\,\ref{tab:notations} are gathered in Table\,\ref{tab:notation2}.  Note
that the optical distances between the beamsplitter and the mirror ${\sf
P}_1$, and between the  beamsplitter and the mirror ${\sf P}_2$ differ by
a quarter of wave length, exactly as in the standard LIGO topology.

The following suppositions and approximation will be used:

\begin{itemize}
  \item The optical $\omega_o$ pumping frequency is much larger than
    all other characteristic frequencies of the system.
	\item The arm cavities are tuned in resonance: $e^{2i\omega_oL/c} = 1$.	
	\item The ``central station'' size is sufficiently small and it is possible 
    to neglect by values of the order of $\dfrac{\Omega l_{\sf D-I}}{c}$,
    $\dfrac{\Omega l_{\sf C-D}}{c}$, $\dfrac{\Omega l_{\sf D-P}}{c}$, and
    $\dfrac{\Omega l_{\sf BS-P}}{c}$.
	\item All optical losses are concentrated in the arm cavities. This assumption
    is reasonable  because losses in arm cavities appear in the
		final expressions amplified by the cavities finesse factor. 	
	\item We neglect the recycling mirrors {\sf P,S} transmittances: $T_{\sf
    S}=T_{\sf P}=0$ because they appear in the final expressions reduced by 
		the mirrors {\sf D} transmittance $T_{\sf P}$.	
	\item Analyzing the power (symmetric) and the signal (anti-symmetric) modes
    we will keep first non-vanishing terms for each mode: classical
    (zeroth-order)  field amplitudes for the power mode and linear in the
		mirror displacements  and in the fields quantum fluctuations ({\it i.e.}
		first-order) terms for the  signal one. 	
\end{itemize}

\subsection{Power mode}

Zeroth approximation equations for the field amplitudes are the following:
\begin{subequations}
  \begin{align}
    {\rm A}_{1,2} &= -R_{\sf D}{\rm D}_{1,2} + iT_{\sf D}{\rm F}_{1,2} \,, \\
    {\rm B}_{1,2} &= R_{\sf FP}(0)A_{1,2}e^{2i\theta'}  \,,
      \label{eqs0_B} \\
    {\rm C}_{1,2} &= -R_{\sf D}{\rm B}_{1,2} + iT_{\sf D}{\rm H}_{1,2} \,, \\
    {\rm D}_{1,2} 
      &= (-R_{\sf C}{\rm C}_{1,2} + iT_{\sf C}{\rm C}_{2,1})e^{2i\theta''}
    \,, \\
    {\rm E}_{1,2} &= -R_{\sf D}{\rm H}_{1,2} + iT_{\sf D}{\rm B}_{1,2} \,, \\
    {\rm F}_{1,2} &= \pm \frac{{\rm E_1}-{\rm E_2}}{2}e^{2i\beta''} 
      + \frac{{\rm F}_0}{\sqrt{2}} \,, \\
    {\rm G}_{1,2} &= -R_{\sf D}{\rm F}_{1,2} + iT_{\sf D}{\rm D}_{1,2} \,, \\
    {\rm H}_{1,2} &= -{\rm G}_{1,2}e^{2i\beta'} \,, \\
    {\rm I}_{1,2} &= \frac{i\sqrt{c\gamma_{\rm load}/L}}{\gamma}\,
      {\rm A}_{1,2}e^{i\theta'} \label{eqs0_I} \,,
  \end{align}
\end{subequations}
where $F_0$ is the input pumping wave amplitude,
\begin{subequations}
  \begin{equation}
    R_{\sf FP}(0) = \frac{\gamma_-}{\gamma}
  \end{equation}
  is  the arm cavities reflection factor at resonance frequency,
  \begin{gather}
    \gamma = \gamma_{\rm load} + \gamma_{\rm loss} \,, \\
    \gamma_- = \gamma_{\rm load} - \gamma_{\rm loss} \,, \\
    \gamma_{\rm load} = \frac{cT_{\sf I}^2}{4L} \,, \\
    \gamma_{\rm loss} = \frac{cA^2}{4L} \,, 
  \end{gather}
\end{subequations}
For Eqs.\,(\ref{eqs0_B}),(\ref{eqs0_I}), see papers \cite{02a2Kh,04a1Da}.

Introduce the symmetric mode (it is easy to see that the anti-symmetric mode
is not pumped):
\begin{equation}
  {\rm A} = \frac{{\rm A}_1 + {\rm A}_2}{\sqrt{2}} \,,
\end{equation}
and correspondingly for all other fields amplitudes. Equations for these
amplitudes are the following:
\begin{subequations}
  \begin{align}
    {\rm A} &= -R_{\sf D}{\rm D} + iT_{\sf D}{\rm F} \,, \\
    {\rm B} &= R_{\sf FP}(0){\rm A}e^{2i\theta'} \,, \\      
    {\rm C} &= -R_{\sf D}{\rm B} + iT_{\sf D}{\rm H} \,, \\
    {\rm D} &= -{\rm C}e^{i(2\theta''-\phi)} \,, \\
    {\rm E} &= -R_{\sf D}{\rm H} + iT_{\sf D}{\rm B} \,, \\
    {\rm F} &= {\rm F}_0 \,, \\
    {\rm G} &= -R_{\sf D}{\rm F} + iT_{\sf D}{\rm D} \,, \\
    {\rm H} &= -{\rm G}e^{2i\beta'} \,, \\
    {\rm I} &= \frac{i\sqrt{c\gamma_{\rm load}/L}}{\gamma}\,
      {\rm A}e^{i\theta'} \label{eqs0_I_symm}\,,
  \end{align}
\end{subequations}
where
\begin{equation}
  \phi = \arctan\frac{T_{\sf C}}{R_{\sf C}} \,.
\end{equation}
Solution of these equations is the following (only those amplitudes which 
will be required later are presented):
\begin{subequations}\label{PL_soln_g}
  \begin{align}
    {\rm A} &= \frac{iT_{\sf D}}{\Det}
      \left[1 + e^{i(2\beta'+2\theta''-\phi)}\right]{\rm F}_0\,,\\
    {\rm C} &= \frac{iT_{\sf D}R_{\sf D}}{\Det}
      \left[e^{2i\beta'} - R_{\sf FP}(0)e^{2i\theta'}\right]{\rm F}_0 
      \,, \\
    {\rm D} &= -\frac{iT_{\sf D}R_{\sf D}}{\Det}\left[
        e^{i(2\beta'+2\theta''-\phi)} - R_{\sf FP}(0)e^{i(2\theta-\phi)}
      \right]{\rm F}_0 \,, \\
    {\rm E} &= -\frac{1}{\Det}\left[
        R_{\sf D}^2e^{2i\beta'} 
	+ R_{\sf FP}(0)
	    \left(T_{\sf D}^2 + e^{i(2\beta'+2\theta''-\phi)}\right)e^{2i\theta'}
      \right]{\rm F}_0 \,, 
  \end{align}  
  where
  \begin{equation}
    \Det = 1 + T_{\sf D}^2e^{i(2\beta'+2\theta''-\phi)}
      + R_{\sf FP}(0)\,R_{\sf D}^2e^{i(2\theta-\phi)} \,.
  \end{equation}
\end{subequations}

Suppose then that cavities ${\sf CI}$ are tuned in resonance and 
cavities ${\sf CP}$ are tuned in anti-resonance:
\begin{subequations}\label{PL_tuning_0}
  \begin{gather}
     e^{i(2\theta-\phi)} = -1 \logequiv 
       2\theta = (\phi + \pi) \hspace{-0.5em}\mod 2\pi \,,\\
     e^{i(2\beta'+2\theta''-\phi)} = 1 \logequiv 
       2\beta' = (-2\theta'' + \phi) \hspace{-0.5em}\mod 2\pi 
       = (2\theta' +\pi) \hspace{-0.5em}\mod 2\pi \,.
  \end{gather}
\end{subequations}
In this case,
\begin{subequations}
  \begin{align}
    {\rm A} &= \frac{iT_{\sf D}\gamma}
      {\gamma_{\rm loss} + T_{\sf D}^2\gamma_{\rm load}}\,{\rm F}_0 \,, \\
    {\rm C} &= -\frac{iR_{\sf D}T_{\sf D}\gamma_{\rm load}}
      {\gamma_{\rm loss} + T_{\sf D}^2\gamma_{\rm load}}\,{\rm F}_0e^{2i\theta'} 
      \,, \\
    {\rm D} &= -\frac{iR_{\sf D}T_{\sf D}\gamma_{\rm load}}
      {\gamma_{\rm loss} + T_{\sf D}^2\gamma_{\rm load}}\,{\rm F}_0 \,, \\
    {\rm E} &= \frac{\gamma_{\rm loss} - T_{\sf D}^2\gamma_{\rm load}}
      {\gamma_{\rm loss} + T_{\sf D}^2\gamma_{\rm load}}\,{\rm F}_0 \,,
  \end{align}
\end{subequations}
If $T_{\rm D}\ll 1$, then the maximum value of the amplitudes ${\rm A,B}$ is
reached when
\begin{equation}\label{T_D_opt}
  T_{\sf D} = \frac{\gamma_{\rm loss}}{\gamma_{\rm load}} \,.
\end{equation}
In this case (we add here Eq.\,(\ref{eqs0_I_symm}) for convenience):
\begin{subequations}\label{PL_soln_0}
  \begin{align}
    {\rm A} &= \frac{i\gamma}{2\sqrt{\gamma_{\rm load}\gamma_{\rm loss}}}\,
      {\rm F}_0 \,, \\
    {\rm C} &= -\frac{iR_{\rm D}}{2}
      \sqrt{\frac{\gamma_{\rm load}}{\gamma_{\rm loss}}}\,{\rm F}_0e^{2i\theta'} 
      = -R_{\rm D}\frac{\gamma_{\rm load}}{\gamma}{\rm A}e^{2i\theta'} \,, \\  
    {\rm D} &= -\frac{iR_{\rm D}}{2}\,
      \sqrt{\frac{\gamma_{\rm load}}{\gamma_{\rm loss}}}\,{\rm F}_0
      = -R_{\rm D}\frac{\gamma_{\rm load}}{\gamma}{\rm A} \,, \\
    {\rm E} &= 0 \quad \text{(there is no reflected wave!)} \,, \\
    {\rm I} &= \frac{i\sqrt{c\gamma_{\rm load}/L}}{\gamma}\,
      {\rm A}e^{i\theta'} \,.
  \end{align}
\end{subequations}

\subsection{Signal mode}

The first-order equations are the following (see papers
\cite{02a2Kh,04a1Da}):
\begin{subequations}
  \begin{align}
    \hat{\rm a}_{1,2}(\omega) &= -R_{\sf D}\hat{\rm d}_{1,2}(\omega)
      + iT_{\sf D}\hat{\rm f}_{1,2}(\omega) \,, \\
    \hat{\rm b}_{1,2}(\omega) 
      &=R_{\sf FP}(\Omega)\hat{\rm a}_{1,2}(\omega)e^{2i\theta'} 
        + \hat{\rm b}_{01,02}(\omega) \,, \\
    \hat{\rm c}_{1,2}(\omega) &= -R_{\sf D}\hat{\rm b}_{1,2}(\omega) 
      + iT_{\sf D}\hat{\rm h}_{1,2}(\omega) \,, \\
    \hat{\rm d}_{1,2}(\omega) &= \left[-R_{\sf C}\hat{\rm c}_{1,2}(\omega) 
      + iT_{\sf C}\hat{\rm c}_{2,1}(\omega)\right]e^{2i\theta''}
      + \hat{\rm d}_{01,02}(\omega) \,, \\
    \hat{\rm e}_{1,2}(\omega) &= -R_{\sf D}\hat{\rm h}_{1,2}(\omega) 
      + iT_{\sf D}\hat{\rm b}_{1,2}(\omega) \,, \\
    \hat{\rm f}_{1,2}(\omega) 
      &=\pm\frac{\hat{\rm e}_1(\omega)-\hat{\rm e}_2(\omega)}{2}e^{2i\beta''} 
        + \hat{\rm f}_0 \,, \\
    \hat{\rm g}_{1,2}(\omega) &= -R_{\rm A}\hat{\rm f}_{1,2}(\omega) 
      + iT_{\sf D}\hat{\rm d}_{1,2}(\omega) \,, \\
    \hat{\rm h}_{1,2}(\omega) &= -\hat{\rm g}_{1,2}(\omega)e^{2i\beta'} \,,\\
    \hat{\rm i}_{1,2}(\omega) 
      &=\frac{i\sqrt{\gamma_{\rm load}/\tau}}{\gamma-i\Omega}\,\left[
          \hat{\rm a}_{1,2}(\omega)e^{i\theta'} + \hat{\rm s}_{1,2}(\omega)
        \right] \,,
  \end{align}
\end{subequations}
where
\begin{subequations}
  \begin{equation}
    \Omega = \omega-\omega_o \,, 
  \end{equation}
  \begin{align}
    \hat{\rm b}_{01,02}(\omega) 
      &=\frac{2\gamma_{\rm load}}{\gamma-i\Omega}\,
          \hat{\rm s}_{1,2}(\Omega)e^{i\theta'} \,, \\
    \hat{\rm d}_{01,02}(\omega) 
		  &=\pm\frac{2i\omega_o{\rm C}_{1,2}R_{\sf C}}{c}\,
			       \hat y(\Omega)e^{2i\theta''} \,, \\
    \hat{\rm s}_{1,2}(\omega) 
     &=-\sqrt{\frac{\gamma_{\rm loss}}{\gamma_{\sf load}}}\,
          \hat{\rm n}_{1,2}(\omega)
          + \frac{\omega_o{\rm I}_{1,2}}{\sqrt{c\gamma_{\rm load}L}}\,
            \hat x_{1,2}(\Omega) \,, 
  \end{align}
	\begin{equation}
    \hat x_{1,2}(\Omega) = \hat x_{{\sf E}\,1,2}(\Omega)
      - \hat x_{{\sf I}\,1,2}(\Omega) \,,
  \end{equation}
  $\hat{\rm n}_{1,2}(\omega)$ are the noises created by the internal losses in
  the Fabry-Perot cavities normalized as zero-point fluctuations and
  \begin{equation}
    R_{\sf FP}(\Omega) = \frac{\gamma_-+i\Omega}{\gamma-i\Omega}
  \end{equation}
\end{subequations}
is the arm cavities reflection factor at the (side-band) frequency $\Omega$.

Introduce the anti-symmetric mode:
\begin{equation}
  \hat{\rm a}(\omega) 
    = \frac{\hat{\rm a}_1(\omega) - \hat{\rm a}_2(\omega)}{\sqrt{2}} \,,
\end{equation}
and correspondingly for all other field amplitudes. Taking into account that
\begin{align}
  {\rm C}_1 &= {\rm C}_2 = \frac{{\rm C}}{\sqrt{2}} \,, &
  {\rm I}_1 &= {\rm I}_2 = \frac{{\rm I}}{\sqrt{2}} \,,
\end{align}
we obtain the following equations for this mode field amplitudes:
\begin{subequations}
  \begin{align}
    \hat{\rm a}(\omega) &= -R_{\sf D}\hat{\rm d}(\omega)
      + iT_{\sf D}\hat{\rm f}(\omega) \,, \\
    \hat{\rm b}(\omega) &= R_{\sf FP}(\Omega)\hat{\rm a}(\omega)e^{2i\theta'} 
      + \hat{\rm b}_0(\omega) \,, \\
    \hat{\rm c}(\omega) &= -R_{\sf D}\hat{\rm b}(\omega) 
      + iT_{\sf D}\hat{\rm h}(\omega) \,, \\
    \hat{\rm d}(\omega) &= -\hat{\rm c}(\omega)e^{i(2\theta''+\phi)} 
      + \hat{\rm d}_0(\omega) \,, \\
    \hat{\rm e}(\omega) &= -R_{\sf D}\hat{\rm h}(\omega) 
      + iT_{\sf D}\hat{\rm b}(\omega) \,, \\
    \hat{\rm f}(\omega) &= \hat{\rm e}(\omega)e^{2i\beta''} \,, \\
    \hat{\rm g}(\omega) &= -R_{\rm A}\hat{\rm f}(\omega) 
      + iT_{\sf D}\hat{\rm d}(\omega) \,, \\
    \hat{\rm h}(\omega) &= -\hat{\rm g}(\omega)e^{2i\beta'} \,, \\
    \hat{\rm i}(\omega) &= 
      \frac{i\sqrt{c\gamma_{\rm load}/L}}{\gamma-i\Omega}\,
      \left[\hat{\rm a}(\omega)e^{i\theta'} + \hat{\rm s}(\omega)\right] \,,
  \end{align}
\end{subequations}
where
\begin{subequations}
  \begin{align}
    \hat{\rm b}_0(\omega) &= \frac{2\gamma_{\rm load}}{\gamma-i\Omega}\,
      \hat{\rm s}(\Omega)e^{i\theta'} \,, \\
    \hat{\rm d}_0(\omega) 
      &= \frac{2i\omega_o{\rm C}R_{\sf C}}{c}\,\hat y(\Omega)e^{2i\theta''}
			\,,\\
    \hat{\rm s}(\omega) &=-\sqrt{\frac{\gamma_{\rm loss}}{\gamma_{\sf load}}}\,
      \hat{\rm n}(\omega)
      + \frac{i\omega_o{\rm A}}{c\gamma L}\,\hat x(\Omega)e^{i\theta'}\,,
  \end{align}
  \begin{equation}
    \hat x(\Omega) = \frac{\hat x_1(\Omega) - \hat x_{2}(\Omega)}{2} \,. 
  \end{equation}
\end{subequations}
Solution of these equations is the following:
\begin{subequations}
  \begin{align}
    \det(\Omega)\hat{\rm a}(\omega) &= -\left[
        R_{\sf D}^2e^{i(2\theta''+\phi)} + T_{\sf D}^2e^{2i\beta''} 
        + e^{i(2\beta+2\theta''+\phi)}
      \right]\hat{\rm b}_0(\omega) 
      - R_{\sf D}\left(1 + e^{2i\beta}\right)\hat{\rm d}_0(\omega) \,, \\
    \det(\Omega)\hat{\rm c}(\omega) 
      &=-R_{\sf D}\left(1+e^{2i\beta}\right)\hat{\rm b}_0(\omega) 
        + \left[
            R_{\sf FP}(\Omega)\left(R_{\sf D}^2 + e^{2i\beta}\right)e^{2i\theta'}
            + T_{\sf D}^2e^{2i\beta'}
          \right]\hat{\rm d}_0(\omega) \,, \\
    \det(\Omega)\hat{\rm d}(\omega) 
      &=R_{\sf D}e^{i(2\theta''+\phi)}\left(1+e^{2i\beta}\right)
          \hat{\rm b}_0(\omega) 
        + \left[
            1 + R_{\sf D}^2e^{2i\beta} 
	    + R_{\sf FP}(\Omega)T_{\sf D}^2e^{2i(\beta''+\theta')} 
          \right]\hat{\rm d}_0(\omega) \,,
  \end{align}
  where
  \begin{equation}
    \det(\Omega) 
      = 1 + R_{\sf D}^2e^{2i\beta} + T_{\sf D}^2e^{i(2\beta'+2\theta''+\phi)}
      + R_{\sf FP}(\Omega)\left[
          R_{\sf D}^2e^{i(2\theta+\phi)} + T_{\sf D}^2e^{2i(\beta''+\theta')} 
	  + e^{i(2\beta+2\theta+\phi)}
        \right] \,.
  \end{equation}  
\end{subequations}

Let now conditions Eqs.\,(\ref{PL_tuning_0}) and, in addition, the dark port 
condition for the {\sf D-P} cavities:
\begin{equation}
  e^{i(2\theta''+\phi)} = e^{2i\beta''} \logequiv 
  2\beta'' = 2\theta'+\phi \hspace{-0.5em}\mod 2\pi
    = -2\theta' + 2\phi + \pi \hspace{-0.5em}\mod 2\pi 
\end{equation}
are fulfilled. In this case,
\begin{equation}
  \det(\Omega) = \left[1 + e^{2i\phi}\right]
    \left[1 - R_{\sf FP}(\Omega)e^{2i\phi}\right] \,,
\end{equation}  
and
\begin{subequations}\label{PL_soln_1}
  \begin{align}
    \hat{\rm c}(\omega) &= \frac{-R_{\sf D}\hat{\rm b}_0(\omega) 
      + R_{\sf D}^2R_{\sf FP}(\Omega)\hat{\rm d}_0(\omega)e^{2i\theta'}}
      {1-R_{\sf FP}(\Omega)e^{2i\phi}} 
      - \frac{T_{\sf D}^2\hat{\rm d}_0(\omega)e^{2i\theta'}}{1+e^{2i\phi}}\,,\\
    \hat{\rm d}(\omega) &= \frac{
        -R_{\sf D}\hat{\rm b}_0(\omega)e^{2i(-\theta'+\phi)}  
        + R_{\sf D}^2\hat{\rm d}_0(\omega)
      }{1-R_{\sf FP}(\Omega)e^{2i\phi}} 
      + \frac{T_{\sf D}^2\hat{\rm d}_0(\omega)}{1+e^{2i\phi}}\,, \\
    \hat{\rm i}(\omega) 
      &=\frac{i\sqrt{c\gamma_{\rm load}/L}}{\gamma-i\Omega}\,
        \frac{
          \left[1+e^{2i\phi}\right]\hat{\rm s}(\omega)
          - R_{\sf D}\hat{\rm d}_0(\omega)e^{i\theta'}
        }{1-R_{\sf FP}(\Omega)e^{2i\phi}} \,.
  \end{align}
\end{subequations}

\subsection{Pondermotive forces}

\subsubsection{Central mirror}

Force acting on the central mirror is equal to (taking into account that 
$\Omega\ll\omega_o$):
\begin{multline}
  \hat F_y(t) = \frac{\hbar\omega_o}{c}\left[
      |{\rm C}_2|^2 + |{\rm D}_2|^2 - |{\rm C}_2|^2 - |{\rm D}_2|^2
    \right] \\
    + \frac{\hbar\omega_o}{c}\left\{
        \intOinfty\left[
            {\rm C}_2^*\hat{\rm c}_2(\omega) + {\rm D}_2^*\hat{\rm d}_2(\omega)
            - {\rm C}_1^*\hat{\rm c}_2(\omega) - {\rm D}_1^*\hat{\rm d}_1(\omega)
          \right]e^{i(\omega_o-\omega)t}\,\frac{d\omega}{2\pi} 
        + \hc
      \right\} \\
  = -\hbar\intOinfty\kappa(\omega)\left[
        {\rm C}^*\hat{\rm c}(\omega) + {\rm D}^*\hat{\rm d}(\omega)
      \right]e^{i(\omega_o-\omega)t}\,\frac{d\omega}{2\pi} + \hc \,,
\end{multline}
where $\hc$ stands for ``hermitian conjugate''.

In the spectral domain this equation has the following form: 
\begin{equation}
  \hat F_y(\Omega) = \hat{\cal F}_y(\Omega) + \hat{\cal F}^+_y(-\Omega) \,,
\end{equation}
where
\begin{equation}
  \hat{\cal F}_y(\Omega) = -\frac{\hbar\omega_o}{c}\left[
    {\rm C}^*\hat{\rm c}(\omega_o+\Omega) 
    + {\rm D}^*\hat{\rm d}(\omega_o+\Omega)
  \right]
\end{equation}
Substituting here field amplitudes (\ref{PL_soln_0}), (\ref{PL_soln_1}) we 
obtain:
\begin{multline}
  \hat{\cal F}_y(\Omega) 
  = \frac{\hbar\omega_o{\rm C}^*R_{\sf D}}
      {c\left[1-R_{\sf FP}(\Omega)e^{2i\phi}\right]}
      \left[
        (1+e^{2i\phi})\hat{\rm b}_0(\omega) 
        - R_{\sf D}[1+R_{\sf FP}(\Omega)]\hat{\rm d}_0(\omega)e^{2i\theta'}
      \right] \\
  = \hat{\cal F}_{y\,\rm loss}(\Omega) + {\cal K}_{yx}(\Omega)\hat x(\Omega)
       - {\cal K}_{yy}(\Omega)\hat y(\Omega) \,,
\end{multline}
where
\begin{subequations}\label{PL_central}
  \begin{align}
    \hat{\cal F}_{y\,\rm loss}(\Omega) &= \frac{
        2\hbar\omega_o{\rm C}^*R_{\rm D}\sqrt{\gamma_{\rm load}\gamma_{\sf loss}}
      }{ic(\Omega_B + i\gamma_{\rm loss} + \Omega)}\, 
      \hat{\rm n}(\omega)e^{i\theta'} \,, \\
    {\cal K}_{yy}(\Omega) 
      &=\frac{2\hbar\omega_o^2|{\rm C}|^2R_{\sf D}^2\gamma_{\sf load}}
          {c^2(\Omega_B + i\gamma_{\rm loss} + \Omega)} \,, \\
    {\cal K}_{yx}(\Omega) 
      &=\frac{2\hbar\omega_o^2|{\rm AC}|R_{\sf D}\gamma_{\sf load}}
          {c\gamma L(\Omega_B + i\gamma_{\rm loss} + \Omega)} \,,
  \end{align}
\end{subequations}
and
\begin{equation}\label{Omega_B_app}
  \Omega_B = \gamma_{\rm load}\tan\phi \,.
\end{equation}
It should be noted that Eq.\,(\ref{Omega_B_app}) is valid only if 
\begin{equation}
  \gamma_{\rm load}\tan\phi \ll c/L \,;
\end{equation}
more precise formula is the following:
\begin{equation}
  \Omega_B 
	  = \frac{c}{L}\arctan\left(\frac{\gamma_{\rm load}L}{c}\tan\phi\right)\,,
\end{equation}
see \cite{02a1Kh}.

\subsubsection{Arm cavities}

Forces which act on the mirrors {\sf I,E} are equal to (see papers 
\cite{02a2Kh,04a1Da}):
\begin{equation}
  \hat F_{x\,1,2}(\Omega) 
  = -\hat F_{\sf I\,1,2}(\Omega) = \hat F_{\sf E\,1,2}(\Omega) 
  = \hat{\cal F}_{x\,1,2}(\Omega) + \hat{\cal F}_{x\,1,2}^+(-\Omega) \,,
\end{equation}
where
\begin{equation}
  \hat{\cal F}_{x\,1,2}(\Omega) 
    = \frac{2\hbar\omega_o{\rm I}_{1,2}^*\hat{\rm i}_{1,2}(\omega)}{c} \,.
\end{equation}
Introduce differential force:
\begin{equation}
  \hat F_x(\Omega) = \hat F_{x\,1}(\Omega) - \hat F_{x\,2}(\Omega) 
  = \hat{\cal F}_x(\Omega) + \hat{\cal F}_x^+(-\Omega)
\end{equation}
For this force we obtain:
\begin{equation}
  \hat{\cal F}_x(\Omega) 
  = \hat{\cal F}_{x\,1}(\Omega) - \hat{\cal F}_{x\,2}(\Omega) 
  = \hat{\cal F}_{x\,\rm fl}(\Omega) - {\cal K}_{xx}(\Omega)\hat x(\Omega)
      + {\cal K}_{xy}(\Omega)\hat y(\Omega) \,,
\end{equation}
where
\begin{subequations}\label{PL_arms}
  \begin{align}
    \hat{\cal F}_{x\,\rm loss}(\Omega) &= \frac{
        2\hbar\omega_o{\sf A}^*\sqrt{\gamma_{\rm load}\gamma_{\rm loss}}
      }{i\gamma L(\Omega_B + i\gamma_{\rm loss} + \Omega)}\,
      \hat{\rm n}(\omega)e^{-i\theta'} \,, \\
    {\cal K}_{xy}(\Omega) &=\frac{
        2\hbar\omega_o^2|{\rm AC}|R_{\sf D}\gamma_{\rm load}
      }{c\gamma L(\Omega_B + i\gamma_{\rm loss} + \Omega)} \,, \\
    {\cal K}_{xx}(\Omega) &= \frac{
        2\hbar\omega_o^2|{\rm A}|^2\gamma_{\rm load}
      }{(\gamma L)^2(\Omega_B + i\gamma_{\rm loss} + \Omega)} \,.
  \end{align}
\end{subequations}

\subsection{Mechanical equations of motion}

It is easy to note that 
\begin{subequations}\label{FK_symm}
  \begin{align}
    \hat{\cal F}_{x\,\rm loss}(\Omega) 
      &= -\digamma\hat{\cal F}_{y\,\rm loss}(\Omega)  \,, \\
    {\cal K}_{xx}(\Omega) &= \digamma{\cal K}_{xy}(\Omega) 
      = \digamma{\cal K}_{yx}(\Omega) = \digamma^2{\cal K}_{yy}(\Omega) \,.
  \end{align}
\end{subequations}
where
\begin{equation}
  \digamma = \frac{c}{\gamma_{\rm load}LR_{\sf D}^2} \gg 1\,.
\end{equation}
Therefore, the fluctuation forces spectral densities and the pondermotive 
rigidities are described by the following equations:
\begin{subequations}\label{SK_symm}
  \begin{align}
    S_{x\,\rm loss}(\Omega) &= \digamma^2S_{y,\rm loss}(\Omega) 
      = \frac{8\hbar\omega_oW\gamma_{\rm loss}}{cL}\,
          \frac{\Omega_B^2+\gamma_{\rm loss}^2+\Omega^2}{|{\cal D}(\Omega)|^2}
			\label{Sloss} \,,\\
    K_{xx}(\Omega) &= \digamma K_{xy}(\Omega) 
		 = \digamma K_{yx}(\Omega) = \digamma^2K_{yy}(\Omega) 
     = \frac{8\omega_oW}{cL}\,\frac{\Omega_B}{{\cal D}(\Omega)} \label{Kpond}\,,
  \end{align}
\end{subequations}
where
\begin{equation}
  {\cal D}(\Omega) = (-i\Omega + \gamma_{\rm loss})^2 + \Omega_B^2 \,,
\end{equation}
and
\begin{equation}
  W=\hbar\omega_o|{\rm I}_{1,2}|^2 = \frac{\hbar\omega_o|{\rm I}|^2}{2}
\end{equation}
is the optical power circulating in the arm cavities.

The ``raw'' set of the mechanical equations for all five test masses is
the following:
\begin{subequations}
  \begin{align}
    M_{\sf I}\frac{d^2\hat x_{{\sf I}\,1,2}(t)}{dt^2} 
		  &= -\hat F_{x\,1,2}(t) \,,\\
    M_{\sf E}\frac{d^2\hat x_{{\sf E}\,1,2}(t)}{dt^2} 
      &= \hat F_{x\,1,2}(t) \pm M_{\sf E}a_{\rm sign}(t) \,, \\
    M_{\sf C}\frac{d^2\hat y(t)}{dt^2} &= \hat F_y(t) + \hat F_{\rm meter}(t) \,,
  \end{align}
\end{subequations}
where 
\begin{equation}
  a_{\rm sign}(t) = \frac{L\ddot h(t)}{2}
\end{equation}
is the signal acceleration,
$h(t)$ is the gravitational-wave signal and $\hat F_{\rm meter}$ is the
meter back-action force.

Excluding mechanical degrees of freedom not coupled with the local mirror 
these five equations can be reduced to the following two ones:
\begin{subequations}
  \begin{align}
    M\frac{d^2\hat x(t)}{dt^2} &= \hat F_x(t) + Ma_{\rm sign}(t) \,, \\
    M_{\sf C}\frac{d^2\hat y(t)}{dt^2} &= \hat F_y(t) + \hat F_{\rm meter}(t)\,.
  \end{align}
\end{subequations}

Insert here pondermotive forces calculated in the previous subsubsection and 
rewrite the equations in spectral representation:
\begin{subequations}
  \begin{align}
    \left[-M\Omega^2 + K_{xx}(\Omega)\right]\hat x(\Omega) 
      &=K_{xy}(\Omega)\hat y(\Omega) + \hat F_{x\,\rm loss}(\Omega) 
        + Ma_{\rm sign}(\Omega) \,, \\
    \left[-M_{\sf C}\Omega^2 + K_{yy}(\Omega)\right]\hat y(\Omega) 
      &=K_{yx}(\Omega)\hat x(\Omega) + \hat F_{y\,\rm loss}(\Omega) 
        + \hat F_{\rm meter}(\Omega) \,.
  \end{align}
\end{subequations}
Taking into account symmetry conditions (\ref{FK_symm}) we obtain:
\begin{multline}\label{y_Omega_loss}
  -m_+\Omega^	2\left[-m_*\Omega^2 + K_{yy}(\Omega)\right]\hat y(\Omega) \\
	 = \mu K_{yy}(\Omega)\digamma a_{\rm sign}(\Omega)
	   - \mu\Omega^2\hat F_{y\,\rm loss}(\Omega) 
   + \left[-\mu\Omega^2 + K_{yy}(\Omega)\right]\hat F_{\rm meter}(\Omega) \,.
\end{multline}

The ratio of the first two terms in the right-hand part of this equation 
defines the sensitivity limitation imposed by optical losses. Spectral
density of the corresponding equivalent noise (normalized as fluctuation 
metrics variation) is equal to: 
\begin{equation}\label{opt_loss}
  S_h^{\rm loss}(\Omega) 
  = \frac{4}{L^2\Omega^4}\,
      \frac{\Omega^4S_y^{\rm loss}(\Omega)}{\digamma^2L^2|K_{yy}(\Omega)|^2}
	= \frac{\hbar c\gamma_{\rm loss}}{2\omega_oWL}\,
	    \frac{\Omega_B^2+\gamma_{\rm loss}^2+\Omega^2}{\Omega_B^2} \,.
\end{equation}

In the next section analyzing the local meter schemes we will neglect optical losses
both in the main (large) scheme and in the local meter. In this case equation 
(\ref{y_Omega_loss}) can be simplified:
\begin{equation}\label{y_Omega}
  D(i\Omega)\hat y(\Omega) 
  = \mu\Omega_B^2\Omega_0^2\digamma a_{\rm sign}(\Omega)
    + D_F(i\Omega)\hat F_{\rm meter}(\Omega) \,, 
\end{equation}
where
\begin{subequations}
  \begin{gather}
    D(s) = m_+s^2(s^4 + \Omega_B^2s^2 + \Omega_B^2\Omega_0^2) \,, \\
		D_F(s) = \frac{\mu}{m_*}\,s^2(s^2+\Omega_B^2) + \Omega_B^2\Omega_0^2
		  \,, \\
  \end{gather}
\end{subequations}
\begin{equation}
  \Omega_0^2 = \frac{K_{yy}(0)}{m_*}
   = \frac{8\omega_oW}{\digamma^2m_*cL\Omega_B} \,. \label{Omega0}
\end{equation}

\section{Analysis of the local meter}

\subsection{Ideal variation measurement}\label{app:KLMTV}

The output signal of the meter which monitors the local mirror position $y$,
can be presented as the following [see Eq.\,(\ref{y_Omega})]:
\begin{equation}
  \tilde y(\Omega) = \hat y(\Omega) + \hat y_{\rm meter}(\Omega)
	= \frac{\mu\Omega_0^2\digamma}{D(i\Omega)}
  		\left[a_{\rm sign}(\Omega) + \hat a_{\rm fluct}(\Omega)\right]\,,
\end{equation}
where
\begin{equation}
  \hat a_{\rm fluct}(\Omega) = \frac{
	  D(i\Omega)\hat y_{\rm meter}(\Omega) 
		+ D_F(i\Omega)\hat F_{\rm meter}(\Omega)
  }{\mu\Omega_0^2\digamma} \,,
\end{equation}
and $\hat y_{\rm meter}$, and $\hat F_{\rm meter}$ are meter noises. If 
the meter cavity is sufficiently short then these  noises spectral densities
are equal to:
\begin{align}
  S_y &= \frac{S_0}{\sin^2\phi_{\rm LO}} \,, &
  S_F &= \frac{\hbar^2}{4S_0} \,, &
  S_{yF} &= \frac{\hbar}{2}\cot\phi_{\rm LO} \,,
\end{align}
and
\begin{equation}\label{S_0}
  S_0 = \frac{\hbar c^2T_{\rm local}^2}{64\omega_ow}
\end{equation}
is the {\it residual} noise of the variation meter.

Spectral density of noise $\hat a_{\rm fluct}(\Omega)$ is equal to:
\begin{equation}
  S_a^{\rm meter} = \frac{1}{(\mu\Omega_0^2\digamma)^2}\left[
	  D^2(i\Omega)S_y + 2D(i\Omega)D_F(i\Omega)S_{yF} +	D_F^2(i\Omega)S_F
	\right] \,.
\end{equation}
It reaches minimum if 
\begin{equation}
  \cot\phi_{\rm LO} = -\frac{\hbar}{2S_0}\,\frac{D_F(i\Omega)}{D(i\Omega)}\,,
\end{equation}
and this minimum is equal to:
\begin{equation}\label{S_var}
  S_a^{\rm meter} \equiv \frac{L^2\Omega^4}{4}\,S_h^{\rm meter}
  = \frac{D^2(i\Omega)}{(\mu\Omega_0^2\digamma)^2}\,S_0  
		= \frac{\left[\Omega^4-\Omega^2\Omega_B^2+\Omega_0^2\Omega_B^2\right]^2}
			  {\Omega_0^4\Omega_B^4}\,
			\frac{m_+^2}{\mu^2}\,\frac{\Omega^4S_0}{\digamma^2} \,.
\end{equation}

\subsection{DSVM-based local meter}\label{app:dsvm}

%\subsubsection{General case}

In the time-domain form equation
(\ref{y_Omega}) can be presented as the following:
\begin{equation}
  {\bf D}\hat y(t) = \mu\Omega_B^2\Omega_0^2\digamma a_{\rm sign}(t)
	  + {\bf D}_F\hat F_{\rm meter}(t) \,,
\end{equation}
where
\begin{align}
  {\bf D} &= D(d/dt) \,, & {\bf D}_F &= D_F(d/dt) \,.
\end{align}
The local meter output signal is equal to:
\begin{equation}
  \tilde y(t) = \hat y(t) + \hat y_{\rm meter}(t) \,,
\end{equation}
where $\hat y_{\rm meter}(t)$ is the meter additive noise.

Following the DSVM procedure (see \cite{00a1DaKhVy}) we suppose that: (i)
noises $\hat y_{\rm meter}(t)$ and $\hat F_{\rm meter}(t)$ correlate with
each other:
\begin{equation}
  \hat y_{\rm meter}(t) = \hat y_{\rm meter}^{(0)}(t) 
	  + \alpha(t)\hat F_{\rm meter}(t)
\end{equation}
where $\alpha(t)$ is some given function, and (ii) during a sufficiently short
time interval $\tau$ the signal $a_{\rm sign}(t)$ can be considered as  
constant one. Estimate for this constant can be found using the following 
equation:
\begin{multline}
  \tilde a_{\rm sign} = \frac{1}{\mu\Omega_B^2\Omega_0^2\digamma\bar v}\,
	  \int_\tau v(t){\bf D}\tilde y(t)\,dt \\
	= a_{\rm sign} + \frac{1}{\mu\Omega_B^2\Omega_0^2\digamma\bar v}
	   \int_\tau v(t)\left\{
		   {\bf D}\hat y_{\rm meter}^{(0)}(t)
			 + \left[{\bf D}\alpha(t) + \frac{1}{m_+}{\bf D}_F\right]
					 \hat F_{\rm meter}(t)
		 \right\}dt			 
\end{multline}
where $v(t)$ is filter function and 
\begin{equation}
  \bar v = \int_\tau v(t)\,dt \,.
\end{equation} 

For the short local meter cavity the local meter noises can be considered as
``white'' or $\delta$-correlated ones. 

The term proportional to the back-action force $\hat F_{\rm meter}(t)$ can be 
canceled by the proper choice of $\alpha(t)$ (and this is the essence of the
variation measurement). In this case the measurement error will be equal to:
\begin{equation}\label{Delta_a}
  (\Delta a)^2 = \left(\frac{1}{\mu\Omega_B^2\Omega_0^2\digamma\bar v}\right)^2
	  S_0\int_\tau\left[{\bf D}v(t)\right]^2\,dt \,,
\end{equation}
where $S_0$ is the residual meter noise $\hat y_{\rm meter}^{(0)}(t)$ spectral
density, see Eq.\,(\ref{S_0}).

Therefore, filter function $v(t)$ that provide minimum to the measurement
error functional should satisfy the following Lagrange equation:
\begin{equation}\label{Lagrange}
  {\bf D}^2v = 1\,,	
\end{equation} 
with the following boundary conditions 
\begin{align}\label{bconds}
  v(\pm\tau/2) &= 0, & \left.\frac{d^nv(t)}{dt^n}\right|_{t=\pm\tau/2} &= 0 
   \qquad (n = 1..5)\,.
\end{align}
Solution of this equation can be represented as the following:
\begin{equation}
  v(t) = \dfrac{t^4}{24m_+^2\Omega_B^4\Omega_0^4} 
	  + C_1t^2 + C_2 + C_3t\sin\Omega_+t + C_4\cos\Omega_+t
		+ C_5t\sin\Omega_-t + C_6\cos\Omega_-t\,,
\end{equation}
where 
\begin{equation}
  \Omega_\pm^2 = \frac{\Omega_B^2}{2}
	  \pm\sqrt{\frac{\Omega_B^4}{4}-\Omega_B^2\Omega_0^2} \,,
\end{equation}  
and coefficients $\{C_i\}$ can be found from boundary conditions
(\ref{bconds}). We do not give the exact formulae for $\{C_i\}$ because they
are quite cumbersome and will add to our article several more pages (not very
informative ones, we think).

Being substituted to (\ref{Delta_a}) function $v$ will give the minimum 
measurement error:
\begin{equation}\label{Delta_a_gen}
  (\Delta a)^2 = \frac{S_0}{(\mu\Omega_B^2\Omega_0^2\digamma)^2\bar v}
	= \frac{720}{\tau^5}\,\frac{m_+^2}{\mu^2}\,\frac{S_y}{\digamma^2}\,
	  	\frac{1}{{\cal G}(\Omega_B,\Omega_0)} \,,
\end{equation}  
where 3d-graphics of the function ${\cal G}(\Omega_B,\Omega_0)$ is presented 
in Fig.\ref{fig:F_3D}. It should be noted that function ${\cal G}$ is defined
only in the area where $\Omega_0<\Omega_B/2$ and the frequencies $\Omega_\pm$ 
are real and the system is dynamically stable. The exact expression for ${\cal
G}$ we do not give in this paper due to the same reason as for coefficients
$\{C_i\}$. 

Sequence of discrete measurements with the measurement error $\Delta a$ is 
equivalent to continuous monitoring of the signal acceleration $a$ with the 
sensitivity defined by the equivalent spectral density 
\begin{equation}\label{S_meter}
  S_a^{\rm meter}	\equiv \frac{L^2\Omega^4}{4}\,S_h^{\rm meter}
	= (\Delta a)^2\tau 
	= \frac{720}{\pi^4}\,\frac{m_+^2}{\mu^2}\,
	    \frac{\Omega_{\rm max}^4S_y}{\digamma^2}\,
			\frac{1}{{\cal G}(\Omega_B,\Omega_0)} \,.
\end{equation} 

%\bibliographystyle{phaip}
%\bibliography{ligo,biblio}

\end{document}